\documentclass[aps,showpacs,superscriptaddress,nofootinbib,preprint,11pt]{revtex4}
\pdfoutput=1
\usepackage{hyperref}
\usepackage{graphicx}
\usepackage{array}
\usepackage{placeins}

\usepackage{amsmath, amssymb, graphics, setspace}
\newcommand{\mathsym}[1]{{}}
\newcommand{\unicode}[1]{{}}

\newcommand{\kinetic}{{\rm{kinetic}}}

\newcommand{\be}{\begin{equation}}
\newcommand{\ee}{\end{equation}}

\newcommand{\rmmax}{{\rm{max}}}

\newcommand{\LDA}{{\rm{LDA}}}

\newcommand{\bfr}{{\bf{r}}}
\newcommand{\bfv}{{\bf{v}}}

\newcommand{\calO}{{\cal{O}}}
\newcommand{\calF}{{\cal{F}}}
\newcommand{\calG}{{\cal{G}}}
\newcommand{\rmc}{{\rm{c}}}
\newcommand{\rmoc}{{\rm{oc}}}

\newcommand {\E}{\nabla\phi}
\newcommand {\bea}{\begin{eqnarray}}
\newcommand {\eea}{\end{eqnarray}}

\def\beq{\begin{eqnarray}}\def\eeq{\end{eqnarray}}
\def\be{\begin{equation}}\def\ee{\end{equation}}

\renewcommand{\vec}[1]{\mathbf{#1}}
\newcommand{\partDer}[2]{\frac{\partial #1}{\partial #2}}

\newcommand{\diverg}{\mathrm{div}}

\usepackage[usenames, dvipsnames]{color}

\preprint{TIFR/TH/16-24}
\begin{document}

\title{The Shear Viscosity in an Anisotropic Unitary Fermi Gas}
\author{Rickmoy Samanta}
\email{rickmoy@theory.tifr.res.in}
\affiliation{Department of Theoretical Physics, Tata Institute of Fundamental Research, Homi Bhabha Road, Colaba, Mumbai 400 005, India}
\author{Rishi Sharma}
\email{rishi@theory.tifr.res.in}
\affiliation{Department of Theoretical Physics, Tata Institute of Fundamental Research, Homi Bhabha Road, Colaba, Mumbai 400 005, India}
\author{Sandip P. Trivedi}
\email{sandip@theory.tifr.res.in}
\affiliation{Department of Theoretical Physics, Tata Institute of Fundamental Research, Homi Bhabha Road, Colaba, Mumbai 400 005, India}
 
\begin{abstract}

We consider a system consisting of a strongly interacting, ultracold unitary Fermi gas under
harmonic confinement. Our analysis suggests the possibility of
experimentally studying, in this system, an anisotropic shear viscosity tensor driven by the
anisotropy in the trapping potential. In particular, we suggest that this 
experimental setup could  mimic some features of anisotropic geometries that
have recently been studied for strongly coupled field theories which have a
gravitational dual. Results using the AdS/CFT correspondence in these theories
show that in systems with a background linear potential, certain viscosity
components can be made much  smaller than the entropy density, parametrically violating the bound proposed by Kovtun, Son and Starinets (KSS). This intuition, along with results from a Boltzmann analysis that
we perform, suggests that a violation of the KSS bound can perhaps occur in the  unitary Fermi  gas system when it is subjected to   a suitable anisotropic trapping potential which may be approximated to be linear in a suitable range of parameters.
 We give a concrete proposal for an experimental setup where an anisotropic
shear viscosity tensor may arise. In such situations, it may also be possible to observe a reduction in the spin one component of the shear viscosity from its lowest value observed so far in ultracold Fermi gases. In
extreme anisotropic situations, the reduction may be enough to reduce the shear
viscosity to entropy ratio  below the proposed KSS bound, although this regime is
difficult to analyze in a theoretically controlled manner.

\end{abstract}
\vskip .5 true cm

\maketitle

\section{Introduction}

The calculation of the transport properties of strongly coupled quantum
theories is a challenging puzzle of interest to theorists working on a wide
range of systems including ultra-cold Fermi
gases at unitarity~\cite{Adams:2012th,Bulgac:2010dg}, heavy ion
collisions~\cite{Adams:2012th,Bhalerao:2010wf}, and neutron
stars~\cite{Page:2006ud,Alford:2014doa}.  

At strong coupling, perturbative expansions fail to give reliable results.
Sophisticated Monte-Carlo techniques which are used to study such theories
non-perturbatively by evaluating path-integrals in imaginary time, while very
successful for calculating equilibrium properties (in the Fermi gas context see
Ref.~\cite{Gezerlis:2007fs} and Refs therein; for heavy ion collisions see
Ref.~\cite{Gavai:2005dy} and Refs therein) can not be easily generalized to
study transport (in the Fermi gas context
see Ref.~\cite{Wlazlowski:2012jb,Wlazlowski:2015yga}; for heavy ion collisions
see Ref.~\cite{Meyer:2009} and Refs therein).

A class of strongly interacting quantum field theories in $d$ dimensions in
some limits can be related to weakly coupled theories of gravity (called their
dual) in $(d+1)$ dimensions. This correspondence~\cite{Maldacena:1999} allows
us to compute dynamical properties of these theories. These computations have
provided many insights into the transport properties of strongly coupled field
theories. 

In certain limits (large t'Hooft coupling $\lambda$ and large number of
``colors'' $N_c$) one can show that for all isotropic theories in $3+1$
dimensions which admit gravity duals, the ratio of shear viscosity $\eta$ to entropy density $s$ is
${\eta \over s} = \frac{1}{4\pi}$~\cite{Son:2002sd,Kovtun:2004de} (we are
working in units with $\hbar=1$ and $k_B=1$). Since weakly coupled theories
typically have much larger ${\eta \over s}$, it was conjectured by Kovtun, Son and Starinets (KSS)  that ${\eta
\over s}$ is bounded from below by $1/(4\pi)$. Subsequently it was found that finite $\lambda$
corrections can drive ${\eta \over s}$ below the KSS
bound~\cite{Brigante:2007nu,Brigante:2008gz,Kats:2007mq,Buchel:2008vz,Sinha:2009ev,Cremonini:2011iq}.

While the theories describing ultra-cold Fermi gases and heavy ion collisions
do not have known gravitational duals and controlled calculations are
difficult, beautiful experiments have managed to measure the value of
$\eta/s$ in the two systems. The value of $\eta/s$ of the quark gluon plasma
created in heavy ion collisions, required for hydrodynamic
simulations to be consistent with the experimentally measured spectrum of
low energy particles (see
Ref.~\cite{Heinz:2013th} for a review), seems to be close to $1/(4\pi)$.
Remarkably, $\eta/s$ has been measured for ultra-cold fermions at unitarity for
a wide range of temperatures and the minimum value (see
Refs.~\cite{Schafer:2007pr,Cao58,Thomas:2015}) is about six times the KSS bound.    

On the other hand the shear viscosity tensor for many interesting systems is
often anisotropic. For example, it has been suggested that the highly
anisotropic initial states in heavy ion collisions (the direction parallel to
the collision axes is fundamentally different from the transverse directions)
may give rise to anisotropic transport properties~\cite{Martinez:2012tu}.
Furthermore, many interesting states of matter, eg. spin density waves and
spatially modulated phases, are anisotropic.  Another possibility, that we
shall explore in detail in this paper, is that an externally applied field can
pick a particular direction and give rise to anisotropies in the shear
viscosity. This possibility has been explored extensively for the case of
weakly coupled theories in the presence of a background magnetic field. (See
Ref.~\cite{Landau1987Fluid} for a classic treatment, Ref.~\cite{Tuchin:2011jw}
for applications to heavy ion collisions and Ref.~\cite{Ofengeim:2015qxz} for
applications to neutron stars.) The behavior of strongly coupled theories in
the presence of an external field is less well explored. With this in mind,
anisotropic gravitational backgrounds in field theory have been recently
studied using the AdS/CFT correspondence (see
\cite{Landsteiner:2007bd,Azeyanagi:2009pr,Natsuume:2010ky,Erdmenger:2010xm,
Basu:2011tt,Erdmenger:2011tj,Mateos:2011ix,Mateos:2011tv,Iizuka:2012wt}) and the behavior of
the  viscosity in  some  of these anisotropic phases has also been analyzed
(see \cite{Rebhan:2011vd, Polchinski:2012nh} and
\cite{Giataganas:2012zy,Mamo:2012sy, Bhattacharyya:2014wfa, Jain:2014vka,Critelli:2014kra,Ge:2014aza}). 

The results of Ref.~\cite{Jain:2014vka} and Ref.~\cite{Jain:2015txa} for example, indicate that one may obtain
parametric violations of the KSS bound in such anisotropic scenarios. This feature arises
in a wide variety of examples and seems to be quite general. In
particular, for a spatially constant driving force which breaks rotational invariance, it was found that by increasing the strength of the driving force compared to the temperature, the ratio for appropriate components of the shear viscosity to entropy density 
can be made arbitrarily small, violating the KSS bound. \\
If this phenomenon also carries over to the unitary Fermi gases, it may be
possible to measure these small viscosities in experiments with trapped
ultra-cold Fermi gases. For this purpose, it is helpful to consider traps which
share the essential features of the systems in Ref.~\cite{Jain:2014vka,Jain:2015txa}
listed at the end of Sec.~\ref{gravity} of this paper. The goal of this paper is to give a
concrete proposal for the trap geometry and parameters where this effect is
likely to be seen.

While typical trap potentials are harmonic,
[quadratic (Eq.~\ref{eq:harmonic_potential}) rather than linear in the
distance] by using existing results for the thermodynamics of unitary Fermi
gases, we show that for a range of temperatures the dominant contribution to the
damping of collective modes due to viscosity arises from a narrow region in the
trap not near the center, where the trapping potential can be approximately
considered as linear.  In analogy with Ref.~\cite{Jain:2014vka,Jain:2015txa} it is
desirable to have traps that are highly anisotropic, which can be simulated by
taking the trapping frequencies~\cite{Ketterle:2008} in one of the directions
(say $\omega_z$) to be much larger than the frequencies in the other
directions. \\
We describe two hydrodynamic modes whose dissipation is governed by the components of viscosity which are expected to become small in the anisotropic situation considered here. One of them is known in the literature as the scissor mode which has been well
studied for bosonic superfluids at $T=0$
theoretically~\cite{PhysRevLett.83.4452} and has also been experimentally
excited in both bosonic~\cite{PhysRevLett.84.2056} and
fermionic~\cite{PhysRevLett.99.150403}  superfluids. The
second mode is a new quasi-stationary solution to the hydrodynamic equations.
Especially for the scissor mode, we show that for experimentally reasonable values of trap parameters,  the damping rate  of the mode  lies within an experimentally accessible range.
It should therefore be possible to  study this mode, measure  the relevant component of the viscosity and its possible suppression.

To gain some additional understanding of how the anisotropic system might behave, we also make a rough estimate of the viscosity components in the presence of an anisotropic trapping 
potential  using the Boltzmann equation. We find that as the anisotropy increases, due to an increase in the trapping frequency $\omega_z$ in one of the directions, some components of the viscosity tensor decrease,
compared to their value in the isotropic case. 

The outline for the paper is as follows. We review the relevant
results~\cite{Jain:2014vka,Jain:2015txa} for anisotropic theories with
gravitational dual in Sec.~\ref{gravity} and summarize the essential features
required in a system to exhibit the suppression of $\eta/s$. Further details on the gravity results is also provided in Appendix \ref{grdetails}.

Next, we consider the unitary Fermi gas in an  anisotropic harmonic trapping
potential and describe the two hydrodynamic modes  which couple to the small
components of the shear viscosity tensor in Sec.~\ref{vprofile}. In
Appendix~\ref{sci} and \ref{stat} we show that these two hydrodynamic modes
satisfy the equations of superfluid hydrodynamics. Sec.~\ref{edisp} discusses
the energy dissipation due to shear viscosity in these two modes we have
studied. In Sec.~\ref{validity} we examine the constraints on the mode
amplitudes by demanding validity of fluid mechanics and in Sec.~\ref{oc} we discuss the damping in the outer regions of the cloud. Next we review the
thermodynamics of the system in Sec.\ref{thermo}. In Sec.~\ref{trap} we give
parameter values for traps (the trapping potential, the temperature and the
chemical potential at the center of the trap) which are tuned such that the
system possesses the required essential features, and show that  by measuring
the damping rate of  fluid modes (described in Sec.~\ref{vprofile}) one can
measure the shear viscosity. This section contains some of the key results in
the paper.  Sec.~\ref{local} discusses an analysis in a weakly coupled
anisotropic theory using the Boltzmann equation.  We conclude our discussion in
Sec.~\ref{res}. 

The solution of the Boltzmann equation used to estimate the values of the trap
potentials for which we expect the corrections to the viscosity to be
substantial is given in Appendix~\ref{microscopic}. In Appendix~\ref{modes} we
compare the modes (discussed in Sec.~\ref{vprofile}) with the well known 
breathing modes.

\section{Results of shear viscosity from gravity}
\label{gravity}

We briefly review results of computations 
of shear viscosity in the gravity picture obtained by studying anisotropic
blackbranes~\cite{Jain:2014vka} where the breaking of isotropy is due to an externally applied
force which is translationally invariant. The simplest system
discussed in Ref.~\cite{Jain:2014vka} consists of a massless dilaton
minimally coupled to gravity, and a cosmological constant. The action is 
\begin{equation}
S = \frac{1}{16\pi G} \int d^5 x \sqrt{g}~ [R+12\Lambda -
\frac{1}{2}\partial_\mu \phi\partial^\mu\phi]\;,~\label{eq:5dlag}
\end{equation}
where $G$ is Newton's constant in $5$ dimensions and $\Lambda$ is a cosmological
constant. The dual field theory in the absence of anisotropy is a $3+1$ dimensional
conformal field theory. The dilaton profile, linear in the spatial co-ordinate
$z$
\begin{equation}
\label{anisoparam}
\phi=\rho z\;,
\end{equation}
explicitly breaks the symmetry to $2+1$.\\
The conservation equation for the stress tensor gets modified as
\begin{equation}
\partial_\mu T^{\mu\nu}=\langle O \rangle \partial^\nu \phi~\label{eq:hydrodynamics}\;,
\end{equation}
where $O$ is the operator dual to the field $\phi$. The right hand side arises
because the varying dilaton results in a driving force on the system.  We see
that a linear profile results in a constant value for $\partial^\nu \phi$ and
thus a constant driving force. 

%
%
Using AdS/CFT one finds~\cite{Jain:2014vka} that for a system at temperature
$T$, (using the compact notation $\eta_{ijij}=\eta_{ij}$) $\eta_{xz}=\eta_{yz}$
(which are spin 1 with respect to the surviving Lorentz symmetry) is affected by
the background dilaton. In the low anisotropy regime ($\rho/T \ll 1$):
\begin{equation}
\frac{\eta_{xz}}{s}=
\frac{1}{4\pi}-\frac{\rho^2 \log 2}{16 \pi^3 T^2}
+
\frac{(6-\pi^2+54 (\log 2)^2)\rho^4}{2304\pi^5 T^4}
+
{\cal{O}}\bigg[\bigg(\frac{\rho}{T}\bigg)^6\bigg]~\label{eq:eta_low}\;.
\end{equation}
The correction to the zero anisotropy result, the KSS bound
$\frac{\eta_{xz}}{s}=\frac{1}{4\pi}$, is proportional to $\frac{(\nabla
\phi)^2}{T^2}$ where $\nabla \phi=\rho\hat{z}$ is the driving force and $1/T$
is the microscopic length scale in the system. \\
In extreme anisotropy ($\rho/T \gg 1$),
\begin{equation}
{\eta_{xz}}/{s}\rightarrow ({1}/{4\pi}) ({32\pi^2 T^2}/{3\rho^2})\,
\end{equation}
and hence becomes parametrically small~\cite{Jain:2014vka}. But this domain
will not be physically accessible in the cold atom systems.\\
In contrast the $\eta_{xy}$ component (which couples to a spin $2$ metric
perturbation) was found to be unchanged from its value in the isotropic case, 
$\frac{\eta_{xy}} {s}= \frac{1}{4\pi}$.\\
Parametric reduction of the spin $1$ components of $\eta/s$ has been found for
a variety of strongly coupled theories with a gravitational
dual~\cite{Jain:2015txa,Rebhan:2011vd}. 
Motivated by the generality of the above results, (see \cite{ Jain:2015txa}) in the gravity side, we may hope to find
parametrically suppressed viscosities compared to the KSS bound in systems
where the following basic requirements are met.
\label{cond}
\begin{enumerate}
\item{The system is strongly interacting and in the absence of anisotropy have
a viscosity close to the KSS bound.}
\item{The equations of hydrodynamics for the system admits modes sensitive to
the spin one viscosity components as described above and in
Ref.~\cite{Jain:2014vka,Jain:2015txa}.}
\item{Sufficient anisotropy needs to be introduced in the system (say in the
$z$ direction with rotational symmetry preserved along the $x-y$ plane), such
that these spin one components of the viscosity, when measured  in units of the
entropy density, show an experimentally  measurable decreasing tendency from its lowest value observed so far in ultracold Fermi gases.}
\item{The force responsible for breaking of isotropy is approximately
spatially constant.}
\item{The velocity gradients are small enough (compared to say the inverse mean free path) 
ensuring that hydrodynamics is the appropriate effective theory to describe the
system.}
\end{enumerate}

In the next section (Sec.~\ref{Fermi}) we explore a system of trapped
ultra-cold Fermi gases, chosen so as to explore anisotropic fluid
dynamics. While some of the details of this system are different from the
systems with dual gravitational theories discussed above, it is possible to
choose a set of parameters such that the system has the five features listed above. It can therefore be used  to explore the behavior of the 
 viscosity in the anisotropic regime. 

While gravitational duals for the ultra-cold Fermi gases are not yet known and
hence we can not calculate the anisotropic viscosity coefficients in this
strongly coupled system, if the main feature that $\eta_{xz}$ is smaller than
the KSS bound holds true for these, one could potentially measure this
phenomenon in experiments.

\section{Anisotropic viscosity in trapped anisotropic Fermi gases}
\label{Fermi}

Trapped ultra-cold Fermi gas with their scattering length tuned to be near the
unitarity limit~\cite{Stringari:2008,Ketterle:2008}, are strongly interacting
systems for which $\eta/s$~\cite{Schafer:2007pr,Cao58,Thomas:2015}, was
measured to be close to the KSS bound $1/(4\pi)$.  In this section we shall
explore the properties of this system, when it is placed in an anisotropic
trap. We identify suitable hydrodynamic modes  which probe the viscosity
component expected to be suppressed due to the potential in a highly
anisotropic harmonic trap and find that for reasonable choices of parameters
the five criterion referred to above, (see Sec.\ref{cond}), can be met in these
modes. This leads us to suggest that an anisotropic shear viscosity can arise in such systems and appropriate components of the viscosity may show a reduction from the isotropic values in an experimentally accessible way.  

One method~\cite{Thomas:2015} to measure the viscosity
is by starting with an initial state where the fluid is trapped in an
anisotropic harmonic trap. On removing the trapping potential, the fluid
experiences elliptic flow and the extent of the flow is related to the initial
anisotropy and the viscosity. The relevant bulk viscosity of the system vanishes~\cite{Son:2007,Taylor:2010}, which allows one to cleanly extract the shear
viscosity. Note that even though the initial state of the fluid is anisotropic,
the experiment does not probe anisotropic shear viscosities: after the trap potential is removed, the viscosity
tensor at any point is isotropic.

An alternative technique is to measure the damping rate of breathing
modes~\cite{Schafer:2007pr,Cao58} which is related to the loss of energy due to
the viscosity. The experiments we propose in this paper use this alternative technique and   propose to measure the relevant component of the shear viscosity  by
measuring the damping of appropriate  hydrodynamic modes.

\begin{figure}
    \begin{tabular}{cc}
        \includegraphics[width=8.5cm]{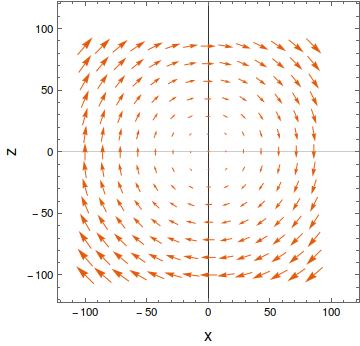}
        &\includegraphics[width=8.5cm]{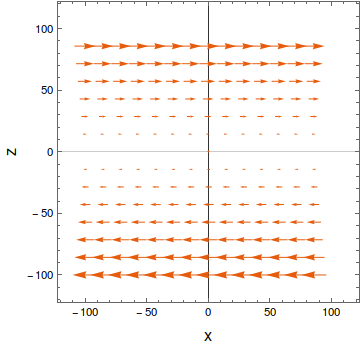}\\
    \end{tabular}
    \caption{(Arbitrary units for coordinates) The flow profile in the $x-z$ plane for {the Elliptic mode}, ie.
    $\bfv=z~\hat{x}-x~\hat{z}$ (left panel, corresponding to
    $\omega_x/\omega_z=1$ in Eq.~\ref{modea}) and
    $\bfv=z~\hat{x}-0.001~x~\hat{z}$ (right panel, corresponding  to
    $\omega_x/\omega_z=0.03$ in Eq.~\ref{modea}).} \label{flow}
\end{figure}

The unitary Fermi gas system we consider here   shares  important features with the
gravitational system described in Sec.~\ref{gravity}. The role of a linear potential was emphasized in Sec.~\ref{gravity}. While such a linear potential cannot arise 
in the trapped fermion system we consider,  we shall
see below  that if we choose the velocity profile and the trap parameters
carefully, the dominant contribution to shear viscosity comes from a region of
the trap where the confining force is approximately constant: satisfying the
fourth criterion listed in Sec.~\ref{gravity}.

The system we consider consists of an ultra-cold Fermi gas under
harmonic confinement described by the potential 
\begin{equation}
\phi(\bfr) =  \sum_i  \frac{1}{2} m \omega_{i}^{2} x_{i}^{2}
~\label{eq:harmonic_potential}
\end{equation}
where $i$ runs over $x,y,z$ and $m$ denotes the mass of the fermionic species.
The trap is anisotropic if $\omega_i$'s are unequal. For example,
$\omega_z\gg\omega_x, \omega_y$ gives rise to a pancake like trap: thin in the
$z$ direction. This can lead to an anisotropic shear viscosity tensor as
described in Sec.~\ref{local}. The potential gradient in the $x$ and $y$
directions is small in most of the trap. 

This section is organized as follows.  After a general  discussion we  describe
the two modes of interest  (referred to as the Elliptic mode and the Scissor mode)
in subsection \ref{vprofile}. The equations of superfluid hydrodynamics are
described in Appendix \ref{superfl}, following which, in Appendix
\ref{sci} and \ref{stat} respectively we show that  the Scissor mode and
the Elliptic mode satisfy these equations. The fluid flow profile in the Elliptic mode is similar to that shown in Fig.~\ref{plates}:  a velocity in the $x$ direction with a gradient in the $z$
direction. The scissor mode is well known  in the literature. In subsection \ref{edisp} we show that the dissipation of energy in the two modes of interest is determined by the relevant components of the viscosity tensor (the spin $1$ components described in the previous section). In Subsection \ref{validity} we find a constraint on the magnitude of the velocity for the two modes by demanding the validity of fluid mechanics. The thermodynamics of the system is discussed in subsection \ref{thermo}.  Finally in subsection \ref{trap} we bring this understanding together and 
show that for reasonable values of parameters the required criterion listed in Sec.~\ref{cond} can indeed be met. 

\begin{figure}
 \includegraphics[width=8.5cm]{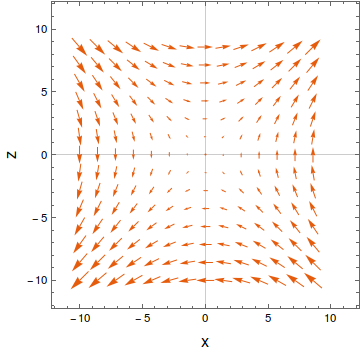}
 \caption{(Arbitrary units for coordinates) The flow profile in the $x-z$ plane
 at time $t=0$ for the Scissor mode,
 ie.  $v=z~\hat{x}+x~\hat{z}$ (Eq.~\ref{scissor})}
\label{scimode}
\end{figure}

\subsection{Choice of Velocity Profile}
\label{vprofile}
Here we first describe the two modes of interest which arise as solutions to
the equations of ideal superfluid hydrodynamics. Each of these modes is
characterized by the superfluid and the normal components, which we denote by
$\textbf{v}_s$ and $\textbf{v}_n$ respectively.\\ The first mode, which we call
the Elliptic mode has  $\textbf{v}_s=0$ and $\textbf{v}_n=\textbf{v}$ given by
\begin{equation}
\label{eq:vprofile}
{\bf{v}} ~=e^{i \omega t} (\alpha_x z ~\hat{x} + \alpha_z
x ~\hat{z})
\end{equation}
with the following relations:
\begin{equation}
\label{modea}
{\textbf{Elliptic mode}}:\;\;\omega=0,~
   \alpha_{z}=- \frac{\omega_{x}^{2}}{\omega_{z}^{2}} \alpha_{x}
\end{equation}

The other mode of interest, denoted by  the Scissor mode, has  $\textbf{v}_s=\textbf{v}_n=\textbf{v}$ given by Eq.~\ref{eq:vprofile} with

\begin{equation}
\label{scissor}
{\textbf{Scissor mode}}:\;\;
\omega=\sqrt{\omega_x^{2}+\omega_z^{2}},~
   \alpha_{z}= \alpha_{x}.
\end{equation}
From the right panel in Fig.~\ref{flow} we see that in the high anisotropy
limit $\omega_{z}\gg\omega_{x}$, $\alpha_{z}\to 0$ for the Elliptic mode, and
hence we recover a flow profile similar to that considered
in~\cite{Jain:2014vka} (shown in Fig.~\ref{plates}); namely a time independent
(in the limit of small viscosity) velocity ($\bfv\propto z\hat{x}$) linearly
increasing with the coordinate in the direction of the gradient of the external
potential ($z$), pointing ($\hat{x}$) in the direction perpendicular to the
gradient of the external potential (neglecting $\omega_x$, $\omega_y$. The gradient is in
the $\hat{z}$ direction). To the best of our knowledge, the Elliptic mode has not
been studied in ultra-cold gas experiments.  The scissors mode
which has been studied extensively (for example see
Refs.~\cite{PhysRevLett.83.4452,PhysRevLett.84.2056, PhysRevLett.99.150403}).

\subsection{Energy dissipation due to viscosity}
\label{edisp}
The energy dissipated due to viscosity is given by
\begin{equation}
\begin{split}
\dot{E}_{\kinetic} =& -\frac{1}{2} \int d^3{\bfr}\, \eta_{ijij}({\bfr})\, 
  \left(\partial_iv_j+\partial_jv_i-\frac{2}{3}\delta_{ij}
      \partial_k v_k \right)^2  
   -\int d^3{\bfr} \, \zeta({\bfr})\, \big( \partial_iv_i\big)^2
   ~\label{eq:dissipation}
\end{split}
\end{equation}
where $\eta_{ijij} \equiv \eta_{ij}$ is the relevant component of the shear viscosity and $\zeta$
is the bulk viscosity. We note that for our chosen velocity profiles, the bulk
viscosity contribution vanishes. Also in the traps we will consider, the temperature T is constant throughout the trap. Hence we also ignored contributions from thermal conductivity.

Thus, 
\begin{equation}
\begin{split}
\dot{E}_{\kinetic} &= 
      - \int d^3{\bfr}\, \eta_{xz}({\bfr}) ~\alpha_{x}^{2} 
      (1-  \frac{\omega_{x}^{2}}{\omega_{z}^{2}})^2
      \label{etaexp}
\end{split}
\end{equation}
is the energy dissipation rate for the Elliptic mode, where we have simply
written $\eta_{xzxz}$ as $\eta_{xz}$.

The energy dissipated per unit cycle for the oscillatory time dependent scissor mode is
\begin{equation}
\begin{split}
\dot{E}_{\kinetic} &= -2\int d^3{\bf{r}}\, \eta_{xz}({\bf{r}}) ~\alpha_{x}^{2} .
 \label{etaosc}
\end{split}
\end{equation}

\subsection{Validity of hydrodynamics}
\label{validity}

One expects that hydrodynamics is a valid description of the system as long as
the viscous correction to the stress tensor is small compared to its value in an
ideal fluid (for eg. see Ref.~\cite{LandauPhysical} or Sec.~$10.3.4$
in Ref.~\cite{zwergerbook}). 

For {the Elliptic mode} the contribution to the stress energy tensor from viscosity is 
\begin{equation}
\eta_{xz}\frac{1}{2}(\alpha_x+\alpha_z)\approx\eta_{xz}\frac{1}{2}(\alpha_x)
\end{equation}
where we have assumed $\omega_{z}\gg \omega_{x,\;y}$ and neglected the
contribution from $\alpha_z$ (see Eq.~\ref{modea}).

For {the Scissor mode} the magnitude of the contribution to the stress energy tensor 
from viscosity is 
\begin{equation}
\eta_{xz}\frac{1}{2}(\alpha_x+\alpha_z)=\eta_{xz}(\alpha_x)
\end{equation}
where we have $\alpha_z=\alpha_x$ for {the Scissor mode}.

At any point $\bfr$, hydrodynamics is expected to be valid if the
viscosity contribution is smaller than the pressure $P(\bfr)$,
\begin{equation}
\alpha_{x} \eta_{xz}({\bfr}) \ll  P ({\bfr}) ~\label{eq:hydro_condition1}\;.
\end{equation}

In the outer edges of the trap the pressure becomes small while $\eta$ tends to
a constant~\cite{PhysRevA.72.043605,Bruun:2006kj,Bluhm:2015raa,PhysRevA.71.033607} and
Eq.~\ref{eq:hydro_condition1} is necessarily violated regardless of how small
$\alpha_x$ is chosen. The contribution of this region to the total energy loss
is typically small however. (Note that the expression Eq.~\ref{eq:dissipation}
can not be used to evaluate the energy loss if Eq.~\ref{eq:hydro_condition1} is
not satisfied~\cite{Bluhm:2015raa}.) What we desire is that hydrodynamics
should be a good theory in the region where the energy loss is substantial.
When we consider specific numerical  values for the  parameters of  the trap in
Subsection \ref{trap}, we will identify a point $\bf {r}_{max}$ close to the
edge of the trap, such that the integral Eq.~\ref{eq:dissipation}  receives
most of its contribution for $r<\bf{r}_{max}$. 

We can then define $\alpha_{x}^{\rm{max}} $ by the condition that
for this amplitude the viscosity contribution to the stress energy tensor is
equal to the pressure at the point $\bfr_{\rmmax}$ 
\begin{equation}
\alpha_{x}^{\rm{max}} =  
\frac{P ({\bfr_{\rmmax}})}{\eta_{xz}({\bfr_{\rmmax}})}.
~\label{eq:hydro_condition}\;
\end{equation}

For $\alpha_{x}<\alpha_{x}^{\rm{max}}$ hydrodynamics is valid in the region of
interest. This constraint limits how large $\alpha_x$ and consequently
$\dot{E}_{\kinetic}$ can be. As long as this dominates over other processes of energy loss
(interaction with the environment) this damping can be measured. In Table.~\ref{ptrap2} in Sec.~\ref{trap} we show this numerical limit for the traps described in
that Section.\\
\subsection{ The outer core}
\label{oc} It has been noted
that a naive application of hydrodynamics at the outer region of the trap where
the density of the atoms is very low leads to an unphysical result. Since the
shear viscosity in the ultra-dilute regime has the form $\eta\sim (mkT)^{3/2}$,
($m$ is the mass, $k$ is the Boltzmann's constant and $T$ is the temperature)
the contribution from the tail (or the outer cloud) is independent of the
density, and hence is divergent [\cite{Bruun:2006kj,Bluhm:2015raa,PhysRevA.72.043605,
PhysRevA.76.045602,PhysRevA.94.043644,
PhysRevA.78.053609,PhysRevLett.116.115301,PhysRevA.71.033607}]. The
unphysical result arises because in the outer part of the trap collisions are
rare and hydrodynamics breaks down. In fact the better approximation in this
region is assuming that atom dynamics in this ultra-dilute region is collisionless and hence does not contribute significantly to damping.

Here we use a simple procedure to take this physics into account. We
only consider traps where the chemical potential at the center is positive and
cutoff the damping contribution from the outer cloud by integrating the
viscosity contribution only from the center of the trap up to
${\bf{r}}_{\rm{max}}$ which is defined as the surface where
$\mu-V({\bf{r}}_{\rm{max}})=T$. We have checked that
changing $r_{max}$ by a little (for example by choosing a slightly larger
$r_{max}^0$ by using the condition $\mu-V(r_{max}^0)=0$) gives similar results
for the damping rates. Similar prescriptions have been followed
previously by \cite{Cao58,Thomas:2015} (see~\cite{PhysRevLett.116.115301} for
an overview).

One can also perform a more careful estimate of the contribution from the
outer cloud. To be concrete, let us consider the scissor mode. We follow the procedure described in
Ref.~\cite{PhysRevA.76.045602} which solves the Boltzmann equation in the
dilute regime, rather than assuming that hydrodynamics is accurate in this
region. Their important result is that for the scissor mode~\footnote{ Let us also note that the scissor mode is excited in the $x-y$ plane in Ref.~\cite{PhysRevA.76.045602}. We have taken care of this fact in our calculations and comparisons.} the energy loss rate in the dilute regime can be written as the integral over 
$\eta$ divided by a suppression factor that increases exponentially as a function of the 
trapping potential. More precisely,
\begin{equation}
\langle\dot{E}_{\rm kinetic}\rangle|_{\rmoc}=
-2\alpha^2\int_{\bfr>\bfr_{\rm{max}}} d^3{ \bfr}
\frac{\eta}{1+\omega^2\tau_\eta^2({\bf r})}, \label{edotoc}
\end{equation}
where in the dilute regime (or the ``classical limit'')
\begin{equation}
\tau_{\eta}({\bf r})=
\frac{4.17}{N\bar{\omega}}\left(\frac{kT}{\hbar\bar{\omega}}\right)^{2}
e^{V({\bf r})/kT}\;,
\end{equation}
and the viscosity $\eta$ is given by
\begin{equation}
\eta=\frac{15}{32\sqrt{\pi}}\frac{(mkT)^{3/2}}{\hbar^2}\;.
\end{equation}

The scissor mode frequency is given by,
\begin{equation}
\omega =\sqrt{\omega_x^2 +\omega_z^2}\;,\label{omega} 
\end{equation}
and the geometric mean $\bar {\omega} = ( \omega_x \omega_y \omega_z)^{1 \over 3}$. 

The integral Eq.~\ref{edotoc} is convergent because of the exponential increase in
the relaxation time $\tau_{\eta}(\bfr)$ even if we take the upper limit of
the integral to $\infty$ but for the numerical evaluation we take the upper
limit of the $x$-integration to be $x_\rmmax+L$, for the $y$-integration to be
$y_\rmmax+L$, and $z$-integration to be $z_\rmmax+L$ with 
$L\gg |\bfr_{\rmmax}|$. 

At the core of the trap hydrodynamics is a good approximation (unless $T\ll
T_c$ where the superfluid phonons can move out of equilibrium). This is a
crucial point because Boltzmann transport is not a valid approximation at the
core where the density of atoms is high. As we explained in the last section, as long as
$\alpha_x=\alpha_z=\alpha<\alpha_x^{\rm{max}}$, hydrodynamics is a good approximation and
the local contribution from the viscosity to the stress energy tensor
\begin{equation}
\begin{split}
\alpha ~ \eta (\bfr) 
\end{split}
\end{equation}
is smaller than the pressure
\begin{equation}
\begin{split}
P(\bfr)
\end{split}
\end{equation}
for $\bfr<\bfr_{\rmmax}$. Therefore, using hydrodynamics to evaluate the
damping contribution from the core, we get
\begin{equation}
\langle\dot{E}_{\rm kinetic}\rangle|_{\rmc}=
-2 \alpha^2\int_{\bfr<\bfr_{\rm{max}}} d^3{ \bfr}~{\eta(\bfr)}\;, \label{edotc}
\end{equation}
where the local value of $\eta(\bfr)$ is calculated using the data for $\eta$ from \cite{Thomas:2015}.  The
integration is performed over $x<x_{\rm{max}}$, $y<y_{\rm{max}}$ and
$z<z_{\rm{max}}$. This approximates the actual ellipsoidal region with a
rectangular shape, but we see that this will not change the results substantially since
the contribution from the outer cloud is small.

\begin{table}
    \begin{center}
        \begin{tabular}{|c|c|c|c|}
           \hline
             $T$
           & $\Gamma_{\rmc}$ ($s^{-1}$)
           & $\Gamma_{\rmoc}$ ($s^{-1}$)\\
          \hline 
          $4 T_c/5$  &$23.03$  &$0.0044$      \\
          $2 T_c/3$  &$18.32$  &$0.00009$      \\
          $4 T_c/7$  &$14.6$  &$2.14\times 10^{-6}$\\
          $T_c/2$  &$11.86$  &$4.69\times10^{-8}$      \\
          \hline
        \end{tabular}
    \end{center}
    \caption{Comparison of contributions to the damping rates for the scissor
    mode from the core [$\Gamma$(c) Eq.~\ref{gammac}],
    and the outer core [$\Gamma$(oc) Eq.~\ref{gammaoc}] for the trap parameters we will explore in our paper.}
    \label{table1}
\end{table}

The amplitude decay rate is given by
\begin{equation}
\Gamma = \frac{|\langle\dot{E}_{\rm kinetic}\rangle|}{2\langle E\rangle}
~\label{eq:defGamma}
\end{equation}
$\langle E\rangle$ is the total mechanical energy averaged over a cycle,
\begin{equation} 
\begin{split}
\langle E\rangle&=\frac{1}{2} \int d^3{r}mn({\bf r}) |v|^2({\bf r})\\
&=\frac{1}{2}m\alpha^2 \int d^3{r}mn({\bf r})(z^2 +x^2)\;,
\label{eng}
\end{split}
\end{equation} 
where $v=\alpha e^{i\sqrt{\omega_x^2+\omega_z^2}t}(z\hat{x}+x\hat{z})$. In Eq.~\ref{gammac}, $\alpha^2$ cancels out and we only need $n(\bfr)$ which is obtained from experiments as explained in detail in Sec.\ref{thermo}.

The damping rate contribution from the outer cloud is given by 
\begin{equation}
\begin{split}
 \Gamma_{\rmoc}&=
 \frac{|\langle\dot{E}_{\rm kinetic}\rangle|_{\rmoc}}{2\langle E\rangle} 
 \label{gammaoc}
\end{split}
\end{equation}
and the contribution from the core is given by
\begin{equation}
\begin{split}
 \Gamma_{\rmc}&=
 \frac{|\langle\dot{E}_{\rm kinetic}\rangle|_{\rmc}}{2\langle E\rangle}\;, 
 \label{gammac}
\end{split}
\end{equation}
and the total damping rate Eq.~\ref{eq:defGamma} is the sum of the two.

In Table.~\ref{table1}, for the representative trap parameters which we will be considering later ( $\omega_z =2 \pi \times 10^4$ rads/s,
$\omega_x=\omega_y= 2\pi \times 385$~rads/s and  $\mu=10\mu$K and $T/T_c$
values as given in the table), we present the comparison of the contribution to
damping from the outer cloud and the core in Table.~\ref{table1}. We see that
the damping contribution from the outer cloud is small, especially for the low
temperatures, justifying our approach.  A direct comparison using our technique (where we cut off the integral for $\dot{E}_{\rm{kinetic}}$ at the point of the trap where hydrodynamics breaks down) can only be made for the lowest
temperature ($T/T_F=0.1$) of Ref.~\cite{PhysRevLett.99.150403}. Our
calculations (using the trap parameters of \cite{PhysRevLett.99.150403})give a damping rate of 250 $s^{-1}$ which agrees with experiments (255  $\pm$ 40 $s^{-1}$,  \cite{PhysRevLett.99.150403}). This is a non-trivial check of our methodology and gives us confidence in our approach in this regime.\\

\begin{figure}
    \begin{tabular}{cc}
        \includegraphics[width=8.5cm]{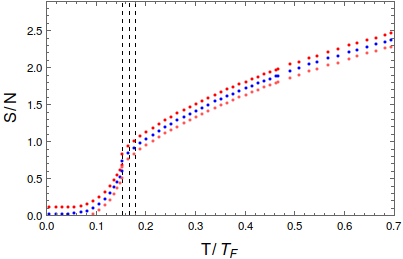}&\includegraphics[width=8.5cm]{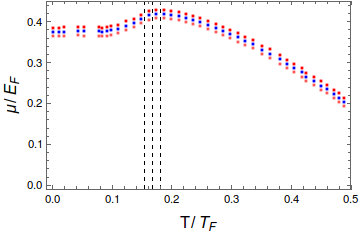}\\
    \end{tabular}
    \caption{(Color online) Data of $\frac{S}{N}$ as a function of $T/T_F$ (left
    panel) and $\mu/E_F$ versus $T/T_F$ (right panel) from Ref.~\cite{Ku563}.
    The central curves (blue online) correspond to the central values and the
    band gives an error estimate (Ref.~\cite{Ku563}). The band denoted by the 
    dashed vertical lines corresponds to the phase transition between the normal and the
    superfluid phase. The error bands represent the maximum error chosen from a set of representative error bars given in Ref.~\cite{Ku563}. }
    \label{sovern}
\end{figure}

\begin{figure}
    \begin{tabular}{cc}
        \includegraphics[width=8.5cm]{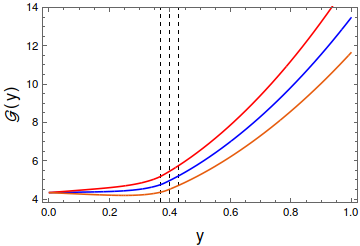}&
        \includegraphics[width=8.5cm]{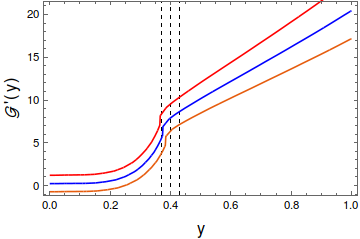}\\
    \end{tabular}
    \begin{center}
        \includegraphics[width=8.5cm]{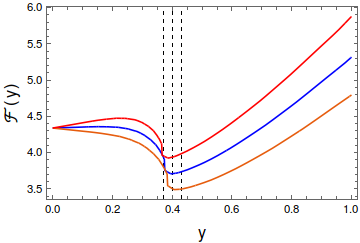}
    \end{center}
    \caption{(Color online) The thermodynamic function $\calG$ (top left panel)
    and its derivative (top right panel) as a function of
    $T\over \mu$. The lower panel shows $\calF$. These dimensionless functions are 
    defined in Eq.~\ref{nands}. The error bands follow from the error bands in Fig.~\ref{sovern}.} 
    \label{gfun}
\end{figure}

\subsection{Thermodynamics}
\label{thermo}

The evaluation of the energy loss from Eq.~\ref{etaexp} and Eq.~\ref{etaosc}
requires the viscosity $\eta$ as a function of the position ${\bfr}$ in the
trap.  In the highly anisotropic traps we are considering the viscosity is
actually a tensor and the different components of the shear viscosity can
acquire different values, in contrast with the isotropic case. For the modes of
interest, Eq.~\ref{eq:vprofile} we need to determine the behavior of  the
component ($\eta_{xz}$).  

To get a first estimate of the region of the trap which gives a dominant
contribution to the integral in Eq.~\ref{eq:dissipation},  we use the local
density approximation (LDA) and  estimate the resulting viscosity.
More specifically, we assume in this approximation that thermodynamic variables like the number density $n$, the entropy density $s$
depend only on  the local value of $T$ and
$\mu$. The  viscosity is also then taken to be given by these local values of $T,\mu$, neglecting any effects of anisotropy which could make the different components of the tensor take different values. 

The effect of anisotropy on the viscosity tensor are estimated using
Eq.~\ref{eq:anisotropy_corrections},  in  a following section
(Sec.~\ref{local}).  While we cannot reliably compute them, the key point of
our analysis here is that they may be experimentally measured and could lie
below the KSS bound. 

To apply  the LDA approximation mentioned above, we start first by considering
a homogeneous system characterized by temperature $T,\;\mu$ and review the
behavior of the thermodynamical parameters and the viscosity as a function of
these parameters. This is covered in this subsection. In the presence of the
trap $\mu$ varies in the equilibrium configuration. The effects of the trap, in
this approximation, are  then incorporated by using the resulting local
value for $\mu$ and $T$ in the behavior for the homogeneous case. The next
subsection will then incorporate the effects of the trap. 
 
In certain thermodynamic regimes, the viscosity of a uniform unitary Fermi gas
can be computed in a controlled manner. At temperatures much smaller than the
chemical potential, transport is dominated by the Goldstone mode associated
with superfluidity and the viscosity can be computed by solving the Boltzmann
transport equations~\cite{PhysRevA.76.053607}. At temperatures large compared
to the chemical potential, the density of fermions is small and a kinetic
estimate of the viscosity, $\eta={\rm{const.}}\times(mT)^{3/2}$, is
adequate~\cite{PhysRevA.72.043605,Bruun:2006kj,Bluhm:2015raa}. But we shall see that
the largest contribution to damping arises from the regime where $T$ and $\mu$
are comparable, and a theoretical evaluation of the viscosity is difficult.
Monte Carlo~\cite{Wlazlowski:2012jb,Wlazlowski:2015yga} methods, microscopic
approaches~\cite{Guo:2010dc}, and $T-$matrix techniques~\cite{Enss:2010qh} have
been used to calculate the viscosity in this regime but presently the best
estimate for the viscosity in this intermediate regime comes from experiments. 

\begin{figure}
\begin{center}
\includegraphics[width=0.45\textwidth]{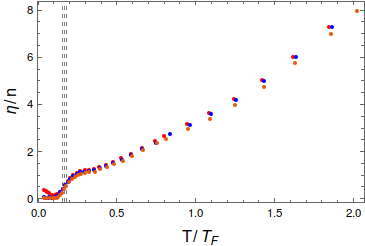}
\includegraphics[width=0.45\textwidth]{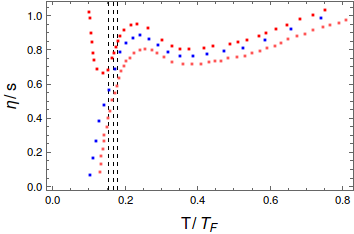}
\caption{(Color online) The left panel shows $\eta \over n$ versus $T/T_F$ from Figs. $2$ and $3$ 
of~\cite{Thomas:2015}. The right panel shows $\eta \over s$ versus $T/T_F$ from Fig. $5$ 
of~\cite{Thomas:2015}.}
\label{etaovern}
\end{center}
\end{figure}

In Refs.~\cite{Schafer:2007pr,Cao58}, $\eta/s$ was measured for the first time.
Recently, this measurement was refined in Ref.~\cite{Thomas:2015} and the result for
the dimensionless ratio $\eta/n$ was measured for a wide range of $T/\mu$,
which we show in Fig.~\ref{etaovern}. Therefore, to obtain the LDA value of the
viscosity, we just need $n(\mu, T)$. 

In the next few paragraphs we describe how to obtain $n(\mu,T)$ using the scaling properties of the unitary Fermi gas. 
With that understanding at hand we will then return to a discussion of how to obtain the viscosity in the approximation described above.
In the unitary Fermi gas, the chemical potential $\mu$ and the temperature $T$
are the only energy scales in the problem. Therefore, we can express various
thermodynamic quantities as a function of the dimensionless quantity $y=T/\mu$
multiplied by an appropriate dimensionful function of only one of the two
variables. Following~\cite{Schafer:2007pr} we write, 
\begin{equation}
\begin{split}
n(\mu,\;T)=&~ n_{f}(\mu)\mathcal{F}(y), \\
s(\mu,\;T)=&~ \frac{2}{5}n_{f}(\mu)\mathcal{G}^{\prime}(y)~\label{nands},
\end{split}
\end{equation}
where $n$ is the number density, $s$ is the entropy density, and
$\mathcal{F}(y)=\mathcal{G}(y)-2~y~\mathcal{G}^{\prime}(y)/5$,
$n_{f}(\mu)={1\over3 \pi^2}(2 m \mu)^{3 \over 2}$ is the number
density of a free Fermi gas. Therefore one can compute the
desired thermodynamic quantities if the function $\mathcal{G}(y)$ is known. For example, one can write the pressure as 
\begin{equation}
\begin{split}
P(\mu,\;T)=&\frac{2}{5} \mu~n_{f}(\mu)~\mathcal{G}(y).
\label{prs}
\end{split}
\end{equation}
In the following discussion, we use the usual definitions 
\begin{equation}
k_F = (3\pi^2 n)^{1/3},\; E_F = \frac{k_F^2}{2m},\;T_F =
E_F/k_B,\;v_F=\frac{k_F}{m}\;. 
\end{equation}

At low temperatures (${T\over T_{F}} \lesssim 0.6$)
we use the $S \over N$ data from Fig.~$3(b)$ of Ref.~\cite{Ku563} to obtain
$\mathcal{G}(y)$. Data from two graphs obtained from Ref.~\cite{Ku563} are shown here in
the two panels of Fig.~\ref{sovern} for convenience. The left panel shows $S/N=s/n$
as a function of $T/T_F$ and the right panel shows $\mu/E_F$ as a function of
$T/T_F$.

In order to solve Eq.~\ref{nands} we need to get $S \over N$ as a function of
$y$. We use Fig. $3(a)$ of Ref.~\cite{Ku563} to convert the  $S \over N$ data in
terms of $y={T\over \mu}$ rather than $T\over T_{F}$. We obtain the function
$\mathcal{G}(y)$ by numerically solving Eq.~\ref{nands}, subject to the
boundary condition $\mathcal{G}(0)=1/\xi^{3/2}$ at  $T=0$. We use $\xi=0.376\pm
0.0075$. (The value of $\xi$ quoted here is from~\cite{Ku563}.  Various
theoretical calculations can be found
in~\cite{Gezerlis:2007fs,Bulgac:2007,Forbes:2010gt,PhysRevB.49.12975,
Goulko:2015rsa, PhysRevLett.96.160402, PhysRevLett.96.090404}.) Fig.~\ref{gfun}
shows the numerically extracted function $\mathcal{G}$ , its first derivative
and the function $\mathcal{F}$.  In Fig.~\ref{gfun} and the rest of the figures, the band
denoted by the dashed vertical lines corresponds to the phase transition
between the normal and the superfluid phase.

The data in Ref.~\cite{Ku563} stops at $T/T_F\approx 0.6$. For higher
temperatures the density is small and as far as thermodynamics is concerned, we
can model the system as a gas of weakly interacting fermions with a self energy
correction in the chemical potential associated with self interactions in the
normal phase. Therefore $n$ and $s$ have the same form as in a Fermi gas, 
(Ref.~\cite{PhysRevA.75.023610})
\begin{equation}
\begin{split}
n_{\rm{norm}} &=
 - g~ (m T)^{3 \over 2}\frac{{\rm{PolyLog}}
 \left(\frac{3}{2},-e^{\mu\over T }\right)}{2 \sqrt{2} \pi ^{3/2}}\\
s_{\rm{norm}} &=
 \frac{\sqrt{T} \left(2~\mu~ {\rm{PolyLog}}
 \left(\frac{3}{2},-e^{\mu \over T }\right)-5~T ~{\rm{PolyLog}}\left(\frac{5}{2},-e^{\mu \over T }\right)\right)}{2 \sqrt{2}\pi ^{3/2}}
\;,~\label{eq:highTthermo}
\end{split}
\end{equation} 
where $n_{\rm{norm}}$, $s_{\rm{norm}}$ denote the number density and entropy in
the normal phase, $g=2$  is the energy level degeneracy, and $\mu$ with self
energy corrections is replaced by $\mu-\frac{3^{2/3} n^{2/3} \pi ^{4/3} (\xi _n
-1)}{2 m}$. Fitting to high temperature data gives $\xi_n \approx
0.45$~\cite{Ku563}. This description works well all the way down to
temperatures $T/T_F\gtrsim0.5$ or equivalently ${T \over \mu} \gtrsim 3.2$ as
one can check by comparing the values of $S/N$ as a function of $T/T_F$ in this
approximation with the results from~\cite{PhysRevA.75.023610}. These results
match smoothly to the low temperature measurements in Ref.~\cite{Ku563}. Therefore
for ${T \over \mu} >3.2$ we use Eq.~\ref{eq:highTthermo} to compute the
thermodynamics.

Now that we have understood how to obtain $n(T,\mu)$ we can return to our discussion of the viscosity. 
To evaluate $\eta$ at a given $\mu$ and $T$ we simply multiply $\eta \over n$
from Fig.~$3$ of Ref.~\cite{Thomas:2015} (shown here in the left panel of
Fig.~\ref{etaovern}) with the number density that can be found using
Eq.~\ref{nands}. One could alternatively multiply $\eta \over s$ from Fig. $5$
of Ref.~\cite{Thomas:2015} (shown here in the right
panel of Fig.~\ref{etaovern}) with the entropy that can be found using
Eq.~\ref{nands}. The former works better because of the smaller error bars. 

As we shall see in the next section when we describe the fermions in a trap,
the dominant contribution to the energy loss arises from the region in the trap
where $T/\mu$ is about $0.54$. This is just above the critical
temperature $T_c$ given by the relation 
\begin{equation}
T_c/T_F = 0.167\pm 0.013\;,~\label{T_cbyT_F}
\end{equation}
or equivalently
\begin{equation}
{T_{c} \over \mu }= 0.4\pm 0.03\;.~\label{T_cbymu}
\end{equation}
From the right panel of Fig.~\ref{etaovern} we see that just above ${T_{c}
\over \mu }\approx 0.4$, $\eta/s\approx 0.7 \approx 8(\frac{1}{4\pi})$. This fact
will be relevant in the next section.

\begin{figure}
\begin{center}
\includegraphics[width=0.45\textwidth]{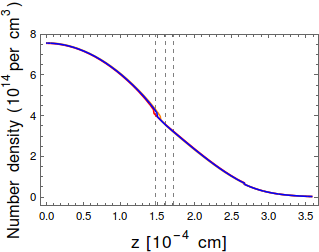}
\includegraphics[width=0.45\textwidth]{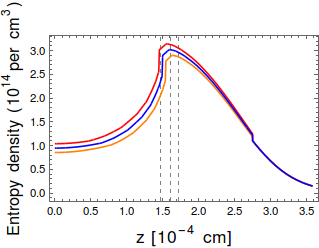}
\caption{(Color online) Variation of number density (left panel) and the
entropy density (right panel) with respect to $z$ for $T={2 T_{c} \over 3}$  at
$\omega_z=2\pi\times10^4$ rads/s with chemical potential at
the trap center 10$\mu$K. The vertical lines denote the band in $z$ where
$T=(0.4\pm 0.03)(\mu-\phi(z))$ (Eq.~\ref{T_cbymu}). }
\label{num}
\end{center}
\end{figure}

\subsection{Results for the trap}
\label{trap}

Having understood the thermodynamics in the absence of the trap, we now turn to incorporating the trap potential in the discussion. 
We first use the LDA approximation to calculate how thermodynamic quantities
like $s, n$ etc. vary along the trap. 
It turns out that on starting at the center of the trap at a sufficiently low temperature,  the entropy density has a peak, $z_0$,  close to the point where the superfluid-normal transition occurs. 
In turn, this  leads to the viscosity and  damping effects  for the fluid modes of interest receiving  their contribution from a region close to the peak and with a  width, $\delta z$ that can be made narrow, $\delta z/z_0<1$. Finally, in this subsection we  examine the resulting behavior of the system for a range of reasonable values of parameters and show that the five conditions listed at the end of Section \ref{cond} can be met. It turns out that both the time scales for energy loss, and the magnitude of the total energy,  lie in the range of experimentally accessible values.

Before we start let us note that there are three energy scales, $T,\mu, \omega_z$ in the system ($\mu$ without an argument refers to the chemical potential at the center of the trap, and we are neglecting $\omega_x, \omega_y$ here). These give rise to  two dimensionless ratios, $T/\mu, \omega_z/\mu$. Length scales can be obtained from these energy scales using the mass, via the relation, $L = {1\over \sqrt{2m E}}$. 

 {\it Thermodynamics in the Trap: }

As discussed in Subsection \ref{superfl} in the presence of a trap the equations for superfluid dynamics can be solved at  equilibrium by taking  the chemical potential to  have a local value
which varies along the trap, as given by \footnote{From now on
$\mu$ without the argument $\bfr$ refers to the chemical potential at the
center of the trap and $\mu(\bfr)=\mu - \phi(\bfr)$.} Eq.~\ref{muans}. The temperature $T$ in equilibrium is a constant. 

Once we have the function $\mathcal{G}$ as discussed in Sec.~\ref{thermo}, one
can then use LDA to express all quantities of interest as a function of the
displacement from the trap center  (which we denote by ${\bfr}$).
Thus, within LDA, we can write the number
density as
\begin{equation}
\label{lda}
 n({\bfr}) = n\left(\mu(\bfr),\;T\right).
\end{equation}
We can also express energy and entropy density in the same fashion as a
function of the distance from the trap center. 
Some comments on  the conditions
for the violation of LDA  will be made in the end of the section.

To set the scales we show (see Fig.~\ref{num}) the number density and the entropy
density as a function of the distance $z$ from the trap center at $x=0, y=0$,
for a typical trap configuration that we consider. In all the examples we
consider, we will take Li$_{6}$ as the fermionic species.  

In making Fig.~\ref{num}, the chemical potential at the center of the trap is
chosen to be $10\mu$K which is typical for experiments performed with fermionic
cold atoms~\cite{Cao58,Caoth}. The potential is taken to be harmonic
(Eq.~\ref{eq:harmonic_potential}), with the confinement frequency along $z$
direction, $\omega_z= 2 \pi \times 10^4$ rads/s which is about 10 times that
chosen in Ref.~\cite{Caoth}.~\footnote{For conversions to energy units, we use
$1$ eV$^{-1}=1.97 \times 10^{-7}$ m, $1$ eV$=1.78 \times 10^{-36}$ kg,
$1$ eV$^{-1}=6.58 \times 10^{-16 }$ s, $1$ eV$= 1.16 \times 10^{4}$ K.  The mass of
Li$_6$ in natural units is $5.6\times 10^{9}$ eV.} Since we are taking $x=y=0$,
$\omega_x$ and $\omega_y$ do not matter in drawing Fig.~\ref{num}. However,
since we will be exploring anisotropic traps we keep in mind  the
condition that $\omega_x=\omega_y\ll\omega_z$. 

The temperature throughout the trap is taken to be
$T=\frac{2 T_{c}}{3}$, where $T_{c}$ is the critical temperature
(Eq.~\ref{T_cbymu}) associated with the chemical potential ($\mu$) at the center
of the trap defined by
\begin{equation}
\label{tcdef}
T_c \equiv 0.4~\mu \;.
\end{equation}
To avoid confusion we note  that $T_c$ is the temperature at which the
superfluid to normal phase transition would have occurred at the center of the
trap.  In the system under consideration with $T=\frac{2T_c}{3}$, since $T$ at
the center of the trap is below the local critical temperature at the center of
the trap, the transition actually occurs away from the center of the trap, at a
location $z=z_c$, where the local chemical potential $\mu(z_c)=\frac{T
}{(0.4)}$ [where we have abbreviated
$\mu((0,0,z_c))$ as $\mu(z_c)$] corresponding to the phase
transition to the normal phase. In Fig.~\ref{num} we
have denoted it by dashed (gray online) vertical lines corresponding to the
central value and the error bands.

The error bands to the densities (marked by red curves online surrounding the blue central curve) are associated
with the errors in $\calG$ (Fig.~\ref{gfun}). They are discontinued from
$z=27.5 \times 10^{-5}$ cm corresponding to the point where we switch to
Eq.~\ref{eq:highTthermo} to calculate the thermodynamics. 

In the other trap geometries we consider below, we will keep the chemical
potential at the center, $\mu$, unchanged as it will set the overall scale of
the problem, and only change the temperature of the trap and the confining
frequency $\omega_z$, in order to explore traps which satisfy
criteria listed in Sec.~\ref{gravity}. The strategy we follow is given below.

As explained in the last section, we estimate the $\eta$ at a given location
${\bfr}$ corresponding to the local chemical potential $\mu(\bfr)$ and 
temperature $T$ by simply multiplying the local number density $n$
we find using Eq.~\ref{nands} with $\eta \over n$ from Fig.~$3$ of
Ref.~\cite{Thomas:2015}. (We have reproduced it here in Fig.~\ref{etaovern} for convenience.) This estimate assumes that not only thermodynamic
but also the transport quantities are determined by the local chemical potential and the
temperature. This estimate necessarily implies that the viscosity is isotropic.
Nonetheless this will help us identify the values of $T/\mu$ for which the energy loss of the
hydrodynamic shear modes is dominated by a region where the potential can be
approximated as a linear potential. Having done that, we will increase
$\omega_z$ to induce anisotropy in the transport coefficients. 

\begin{figure}
    \begin{tabular}{cc}
        \includegraphics[width=9cm]{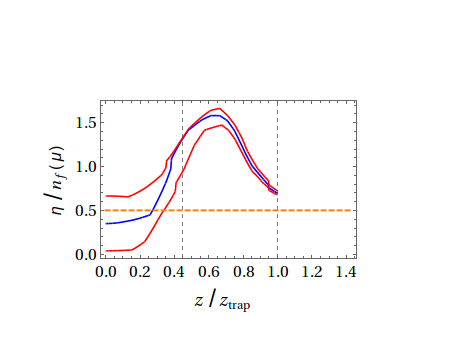}&\includegraphics[width=9cm]{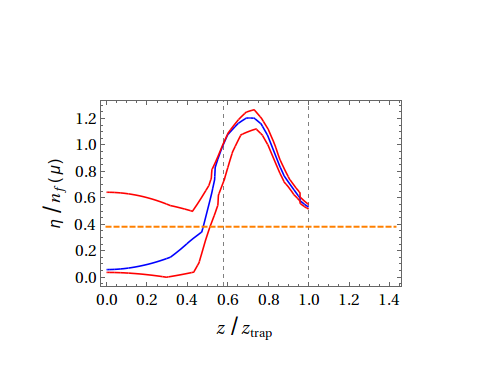}\\
        \includegraphics[width=9cm]{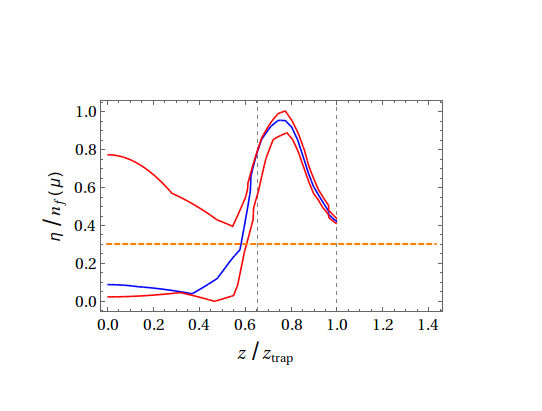}&\includegraphics[width=9cm]{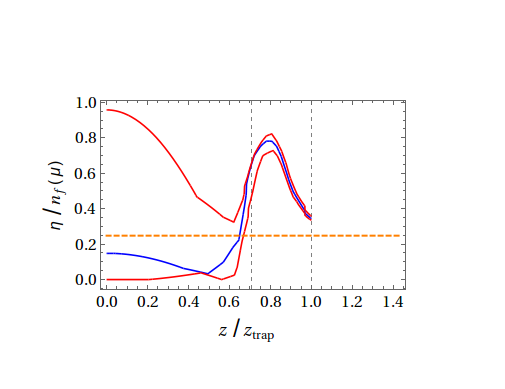}\\
    \end{tabular}
    \caption{(Color online) Local shear viscosity with respect to $z$ for $T={4T_{c} \over 5}$ (top left), $T={2T_{c}
    \over 3}$ (top right) $T={4 T_{c} \over 7}$ (bottom left) and  $T={T_{c} \over 2}$
    (bottom right) at $\omega_z=2 \pi \times 10^4$ rads/s and $\mu=10\mu$K. The red curves around the central blue curve denote the error
    estimate which include errors in the measurement of
    $\eta/n$~\cite{Thomas:2015} as well as errors in $\calG$ due to errors in
    the measurements of thermodynamics~\cite{Ku563}.  The black dashed vertical
    line to the left is at $z_c$ while the one to the right is at $z_{trap}=\sqrt{\frac{2 \mu}{m \omega_z^2}}$. We do not extend the viscosity curves in the dilute regime as discussed in Section \ref {validity}. The  dashed orange horizontal line corresponds to  $\eta/n_f$ in the $\mu \rightarrow - \infty$ limit (\cite{Bruun:2006kj}).}
    \label{localvis}
\end{figure}

Let us consider the four panels in Fig.~\ref{localvis}. They show the local
shear viscosity (in units of $(2m\mu)^{3/2}/(3\pi^2)$ where $\mu$ is the
central chemical potential) as a function of $z$ for $x=0, y=0$ for four
different temperatures at $\omega_z=2 \pi \times 10^4$ rads/s. The chemical
potential at the center is taken to be $10$$\mu$K. The temperatures are
$T={4T_{c} \over 5}$ (top left panel), $T={2T_{c} \over 3}$ (top right panel)
and  $T={4 T_{c} \over 7}$ (bottom left panel) and  $T={T_{c} \over 2}$
(bottom right panel). Like Fig.~\ref{num}, the vertical line (gray online)
corresponds to $z_c$ where $T=0.4\mu(z_c)$. The error bands of the curves are
associated with the errors in ${\cal{G}}$ --- which impact $n$ --- as well as
the errors in the measured $\eta/n$. The $x$-axes of the plots is the $z$
coordinate scaled by the trap size
\begin{equation}
z_{\rm{trap}} = \sqrt{\frac{2  \mu}{m \omega_z^2}}\;.
\end{equation}
One can also define a characteristic distance $z_{\rmmax}$ where $T/\mu(z)=1$ given by 
\begin{equation}
\label{rmax}
z_{\rmmax}= \sqrt{\frac{2(\mu-T)}{m \omega_z^2}}\,.
\end{equation}
For $\mu =10\mu$K at the center of the trap and $\omega_z=2 \pi \times
10^4$ rads/s, $z_{\rm{trap}}$ and $z_\rmmax$ are $\sim
 10^{-4}$ cm. Beyond the distance $z_{\rm{trap}}$, we assume the viscosity to
behave like $\frac{15}{32 \sqrt {\pi}}(m T)^\frac{3}{2}$ as predicted by the
two-body Boltzmann equation~\cite{Bruun:2006kj}.

Note that within LDA the plots in Fig.~\ref{localvis} are independent of
$\omega_z$ if we keep $T/T_c$ fixed. This is because scaling $\omega_z$ by a
factor $f$ can be undone by scaling $z$ by a factor $1/{f}$. Since
$z_{\rm{trap}}$ is scaled by the same factor, $z/z_{\rm{trap}}$ at any point on
the curve remains unchanged.

To understand the behavior of viscosity along the trap, first consider the 
central values in Fig.~\ref{localvis} (blue curve online). For all temperatures
given above (notice that they are all below $T_c$ meaning that the centre of
the trap is superfluid), we find the presence of a peak in the middle region of
the trap length. Qualitatively we understand this from the fact that the local
entropy (see Eq.~\ref{nands}) is the product of $n_f(\mu(\bfr))$ which decreases
along the length of the trap, while the function $\mathcal{G}^{\prime}$
increases along the length of the trap, hence it is natural to expect a peak
for the entropy density somewhere along the length of the trap. It is clearly
seen in the right panel of Fig.~\ref{num}. Since the local shear viscosity over
entropy density is relatively slowly varying in this region (the peak location
is just above the critical region), it is not surprising that the local shear
viscosity shows a similar behavior. Henceforth, we will denote the position
of this peak by $z_{0}$. We also denote the full width at half maximum of the
peak by $\delta z$.  

 The existence of the peak allows us to construct a system where the dominant
contribution comes from a region where the potential approximately varies
linearly, modeling the theories (Sec.~\ref{gravity}) where the force that
breaks rotational invariance is spatially constant. Here, the trap potential
is harmonic, but the dominant contribution to the integral in Eq.~\ref{etaexp} and Eq.~\ref{etaosc}
comes from an interval $\delta z$ near $z_0$. If we expand the confinement
potential as a Taylor series around $z_{0}$ as  
\begin{equation}
 \phi(z_{0}) +  \phi^{'}(z_{0}) (\delta z) +{1 \over 2} \phi^{''}(z_{0})(\delta z)^2 + ......
\end{equation}
The linearity approximation will hold as long as the confinement potential satisfies 
\begin{equation}
\frac{\phi^{''}(z)}{\phi^{'}(z)} ~\delta z \ll 1 \Rightarrow l \equiv 
\frac{ \delta z}{z_{0}} \ll 1\;.
\label{eq:lin}
\end{equation}
Since we are using a harmonic trap, there are no higher order terms. Our
criterion for constant driving force is therefore straightforward. We desire that the
dimensionless ratio $l\equiv{\delta z \over z_{0}}$ be less than $1$.

There are other motivations to choose the dominant contribution to shear
viscosity to arise from such a localized region.  We are interested in
extracting the value of $\eta/s$, for suitable components of the viscosity
tensor, for particular values of $T, \mu$ (in particular, close to the critical
temperature $T_c$ where $\eta/s$ is known to be close to the KSS bound). Due to
the varying trap potential, $\mu(z)$ and therefore  the entropy density at
equilibrium also vary along the trap.  The  change resulting  in the viscosity
due to anisotropy should be  bigger than the effect due to the variation of the
trap potential on $s$, thereby giving rise to the condition,
\begin{equation}
\label{condaa}
{\delta \eta\over \eta}> {\partial s\over \partial z}{\delta z \over s}\,.
\end{equation}
As we saw in Sec.~\ref{gravity} after Eq.~\ref{eq:eta_low} the corrections to
the viscosity due to anisotropy go like square of the force that generates the
anisotropy. For the system at hand this leads to the expectation 
\begin{equation}
\frac{\delta \eta}{\eta} \sim \frac{(\nabla \phi )^2 }{\left(\mu(z)^2
k_F(z)^2\right)}.\label{eq:anisotropy_corrections_rough}
\end{equation}
This estimate agrees with the analysis based on the Boltzmann equation as
discussed later in Sec.\ref{local} (see Eq.~\ref{eq:anisotropy_corrections}).
The RHS in Eq.~\ref{condaa} goes like ${\partial s\over \partial z}{\delta z
\over s}\sim \delta z /z_0=l$,  and this gives rise to the condition
\begin{equation}
\label{conab}
\kappa_{{\rm{LDA}}}^2 >l
\end{equation}
where we have introduced the notation
\begin{equation}
\label{defkappalda}
\kappa_{{\rm{LDA}}}= \frac{(\nabla \phi ) }{\left(\mu(z_0)~ k_F(z_0)\right)} \,.
\end{equation}
It is easy to see that $\kappa_{{\rm{LDA}}}$ roughly scales as
\begin{equation}
\label{condac}
\kappa_{{\rm{LDA}}}\sim\frac{\omega_z}{ \mu}
\end{equation}
so that Eq.~\ref{conab} leads to the condition 
\begin{equation}
\label{conddd}
{\omega_z^2 \over \mu^2}>l.
\end{equation}
For fixed $T,\mu$ one can show that $l$ does not change as  $\omega_z$ changes. Thus the left hand side is independent of the ratio  ${\omega_z \over \mu}$ for fixed $T/\mu$, 
and the inequality can be met for sufficiently large $\frac{\omega_z}{\mu}$. 

Let us also mention that the gravity results apply to situations with only
linearly varying potential (Eq.~\ref{anisoparam}) leading to only
$|\nabla\phi|^2 $ corrections due to the anisotropy. In general we would expect
that there are additional corrections  proportional to $\nabla^2\phi$. There is
little guidance on what these corrections do, for  the kind of strongly coupled
system we are dealing with here.  Thus, to the extent we are trying to stay
close to situations where gravitational systems  give at least some guidance,
it is  desirable to choose the dominant contribution to shear viscosity to
arise from  a narrow localized region.

{\it Viscosity and Other Properties For Varying Trap Parameters: Table \ref{ptrap1} }\\
We now turn to examining the behavior of $\eta$, $\eta/s$, and $l=\frac{\delta
z}{z_0}$ as trap parameters are varied. 
In Table \ref{ptrap1} we keep $\omega, \mu$ fixed to take the values $\omega_z =2 \pi
 \times 10^4$ rads/s,  $\mu=10\mu$K and vary $T$. As mentioned at the beginning of Subsection \ref{trap}  there are two dimensionless ratios that characterize the energy scales in this system.
The different rows  corresponding to different  values of $T$ in units of $T_c$ show how various quantities vary with  $T/\mu$. 
The scaling of these quantities with $\omega_z/\mu$ is given in the first line on top of the Table.~\ref{ptrap1}. Thus $\kappa_{{\rm{LDA}}}$ scales like $\omega_z/\mu$.
$z_0, z_{\rm{trap}}$ and $ \delta z$ scale like $1/\omega_z$ for fixed $T,\mu$,
as was discussed above after Eq.~\ref{rmax}. Thus their ratios, ${z_o\over
z_{\rm{trap}}}, l= {\delta z \over z_0}$ etc. are independent of $\omega_z/\mu$. 
 The third column of the Table.~\ref{ptrap1} tests the linearity of the potential, which is a good approximation near the peak if 
$l=\delta z/z_0 \ll 1$. 

The ratio $l$ is governed by the temperature of the trap divided by the
chemical potential or equivalently $T_c$ at the center. As
we decrease $T/T_c$, $z_0$ increases and $\delta z$ decreases. This
consideration would suggest that to obtain ${\delta z \over z_{0}}$ as small as
possible we should consider as small a temperature as possible. But this
conclusion is not correct as is clear from the upper error band in
Fig.~\ref{localvis} (red online). 
\begin{table}
    \begin{center}
        \begin{tabular}{|c|c|c|c|c|c|c|c|}
            \hline
             $T$          
           & $z_{\rm{trap}}\sqrt{\frac{\mu}{10 \mu {\rm{K}}}}\frac{2 \pi \times 10^4}{\omega}$ cm
           & $\frac{z_{0}}{z_{\rm{trap}}} $
           & $l$
           & $\frac{T}{\mu(z)}|_{z_0}$
           & $\frac{\eta}{n}|_{z_0}$
           & $\frac{\eta}{s}|_{z_0}$
           & $\kappa_{{\rm{LDA}}}\frac{10 \mu {\rm{K}}}{\mu}\frac{\omega_z}{2 \pi \times 10^4 rad/s}$\\
           \hline 
           $4T_{c}/5$    &$27 \times 10^{-5}$   &0.63 &0.98  &0.54 & 0.89 &0.85 &0.05 \\
           $2T_{c}/3$  &$27 \times 10^{-5}$  &0.71 &0.62   &0.54 & 0.89   &0.85 &0.08 \\
           $4 T_{c}/7$  &$27 \times 10^{-5}$   &0.76 &0.46  &0.54 & 0.89   &0.85 &0.11 \\
           $T_{c}/2$   &$27 \times 10^{-5}$   &0.8 &0.37   &0.55 & 0.91   &0.85 & 0.13\\
           \hline
        \end{tabular}
    \end{center}
    \caption{Trap characteristics for various $T/T_c$. The scaling behavior of
    various quantities with $\omega_z$ are also shown. The entries  were
    calculated for $\mu=10\mu$K, $T_c=0.4\mu$. $l=\frac{\delta z }{z_0}$ (Eq.~\ref{eq:lin}) tests
    how well the potential can be approximated as a linear potential in the
    regime of interest. $\kappa_{\rm{LDA}}$ (Eq.~\ref{tf}) tests how well
    LDA is expected to work at $z_0$.}
    \label{ptrap1}
\end{table}

\begin{table}
    \begin{center}
        \begin{tabular}{|c|c|c|c|c|c|c|c|}
             \hline
             $T$          
           & $\alpha_x^{\rm{max}}$($10^{-10}$eV) 
           & ${\dot{E}_{\kinetic}}$(j/s)(\textbf{a})
           & $E$(j) (\textbf{a}) 
           & $\tau_0(s)$(\textbf{a})
           & ${\dot{E}_{\kinetic}}$(j/s)(\textbf{b})
           & $E$(j) (\textbf{b}) 
           & $\tau_0(s)$(\textbf{b})\\
           \hline 
           $ 4T_{c}/5$    &$2.83$   &$2.37 \times 10^{-16}$  &$3\times 10^{-20}$    &$0.0002$ & $4.7 \times 10^{-16}$ & $10^{-17}$ &0.04 \\
           $2T_{c}/3$     &$2.35$    &$1.25 \times 10^{-16}$ &$2\times10^{-20}$     &0.0003 & $2.5\times 10^{-16}$ &6.8 $\times 10^{-18}$ &0.05 \\
           $4 T_{c}/7$    &$2.02$   &$7.12 \times 10^{-17}$ &$1.4 \times 10^{-20}$  &0.0004 & $1.4\times 10^{-16}$ &4.8 $\times 10^{-18}$ &0.07 \\
           $T_{c}/2$      &$1.77$    &$4.33\times10^{-17}$    & $1.1\times 10^{-20}$   &0.0005 &$8.65\times 10^{-17}$  &3.6 $\times 10^{-18}$ &0.08\\
           \hline
        \end{tabular}
    \end{center}
    \caption{Additional trap characteristics for various $T/T_c$ at  $\omega_z =2 \pi
    \times 10^4$ rads/s, $\omega_x=\omega_y= 2\pi \times 385$ rads/s and  $\mu=10\mu$K. The energy is given in joules abbreviated as `j' and energy loss rate in joules per second, (j/s). For a fixed $ T/ \mu$,  the energy
    of the Elliptic mode scales as $ \sim \frac{1}{\omega_x \omega_y \omega_z^3}
    $ and that of the Scissor mode scales as $\sim \frac{1} {\omega_x^3 \omega_y
    \omega_z}$.  The characteristic time $\tau_0$ ( given in seconds `s' in the table and defined in Eq.\ref{taudef}) of  the Elliptic mode scales as $
    \sim \frac{\mu}{ \omega_z^2} $ and that of the Scissor mode scales as $\sim
    \frac{\mu} {\omega_x^2 }$. For the Elliptic mode  to account for the fact that only the normal component of the velocity is non-zero near the trap
    centre, we assume that the normal component density in this region is
    $\frac{T}{T_c}$ times the total density in this region. For the Scissor mode we have the full number density.}
    \label{ptrap2}
\end{table}

The errors bands on $\eta$ are fairly narrow in the region near $z_0$. However,
the errors grow near $z\rightarrow 0$, in particular for smaller $T/T_c$
(Fig.~\ref{localvis}). The reason is the large errors in the measured $\eta/n$
in the superfluid regime (see the region $T/T_F\lesssim0.16$ in
Fig.~\ref{etaovern}). Indeed, we expect that for $T\ll T_F$, the viscosity is
dominated by superfluid phonons whose contribution diverges as $T\rightarrow0$ as
$\eta\approx (9.3\times 10^{-6})~\xi^5~(T_F^8/v^3T^5)$ where $v$ is the speed of
superfluid phonons \cite{PhysRevA.76.053607}. Numerically, $\eta/n\approx
2.5 \times 10^{-5}~ \frac{T_F^5}{T^5}$. Therefore, to avoid a large contribution from the
center of the trap rather than from near $z_0$, we do not consider temperatures
below $T_c/2$. Within this constrained temperature regime between
$T_c/2$ and $T_c$ we find that the linearity condition $\delta z/z_0<1$ is
satisfied, although it is not possible to generate traps where $\delta z/z_0$
is parametrically small. In the narrow range of temperatures, it turns out that
the location of $z_0$ is such that $T/\mu(z_0)\approx0.54$, just off to the
right of the phase transition at $T/\mu(z_c)\approx0.4$. 

Note that, as explained in the discussion above, a few paragraphs after
Eq.~\ref{tcdef},  the value for the viscosity $\eta/s$ which appears in the
Table \ref{ptrap1} is an approximate one, obtained by taking the value in the
isotropic situation corresponding to the local value for $\mu$, $T$ at the
location $z_0$. By a similar argument as before, this value is independent of
the ratio $\omega_z/\mu$ for a fixed $T/T_c$.  We note that the values of
$\eta/s$ in the Table \ref{ptrap1} are about $10$ times the KSS bound. One
would expect that various components of the viscosity tensor deviate from this
rough value by a fraction of order $\kappa_{{\rm{LDA}}}^2$. The parameter
$\kappa_{{\rm{LDA}}}$ which was introduced in Eq.~\ref{defkappalda} above, when
computed at the location of the peak $z_0$, has the more exact form  
\begin{equation}
\label{kappaexact}
 \kappa_{{\rm{LDA}}} =\frac{ m\omega_z^2 z_0}{(3 \pi^2 n(z_0))^{1\over3} \mu(z_0) }= \frac{\sqrt{m \over 2} \omega_z^{2} z_0}
 {[\calF(\frac{T}{\mu(z_0)})]^{1/3}[\mu(z_0)]^{\frac{3}{2}}}
 \end{equation}
as one can easily check by using Eq.~\ref{nands}.

{\it Energy Damping For Varying Values of Trap Parameters: Table \ref{ptrap2}}

We now turn to considering the effects of varying the  trap parameters on various quantities like the total energy $E_{\kinetic}$, the  damping rate of this energy $\dot{E}_{\kinetic}$, etc. 
In Table \ref{ptrap2} we again keep $\mu, \omega_z$ fixed  to take values $\omega_z =2 \pi
    \times 10^4$ rads/s,  $\mu=10\mu$K and consider the effects of varying $T$.  In addition, we also need to consider the effects of the  harmonic trap in the $x,y$ directions. We keep $\omega_x,\omega_y$ to be fixed to take values $\omega_x=\omega_y= 2\pi \times 385$ rads/s. The different rows then give how various  quantities vary as $T/\mu$ changes. 
We note that for the range of temperatures considered the total number of atoms in the trap is approximately, $\sim 10^6$.

The energy which appears in this Table is the total mechanical energy $E$ given by
\beq
E = 2 E_{\rm{kinetic}}
\eeq
where
\begin{equation}
\begin{split}
{E}_{\kinetic} &= 
      \langle \frac{1}{2} \int d^3{\bfr}\, m n({\bfr}) \bfv^2\rangle\;,
      \label{energy}
\end{split}
\end{equation}
where $\bfv$ is the velocity of either mode and the average is taken over one
cycle for the scissor mode (the elliptic mode is non-oscillatory). For the Elliptic mode and the Scissor mode with amplitude
$\alpha_x^{\max}$, the kinetic energy is given as follows: 
\begin{equation}
\label{ekin}
\begin{split}
{\rm{For }}\; {\bf{ Elliptic}},\;
E_{\kinetic}(\textbf{a}) &= \int d^3\bfr \; \frac{1}{2}m\;n_{normal}\;(\alpha_x^{max})^2[\frac{\omega_x^4}{\omega_z^4}x^2+z^2]\\
{\rm{For }}\; {\bf{Scissor}},\;
E_{\kinetic}(\textbf{b}) &= \int d^3\bfr \; \frac{1}{4}m\;n\;(\alpha_x^{max})^2[x^2+z^2]\,.
\end{split}
\end{equation} 
$\dot{E}_{\kinetic}$ is the rate of energy loss due to viscosity induced dissipation, Eq.~\ref{eq:dissipation}. The energy loss, $\dot{E}_{\kinetic}$ in these modes is given by Eqns.~\ref{etaexp}, \ref{etaosc}. 

Note that  for the Scissor mode the expression corresponds to the kinetic energy
averaged over an oscillation cycle. Also, for the Elliptic mode, $v_s=0$, Eq.~\ref{modea},  and only the normal component contributes to the kinetic energy.  
The density in the  normal phase is estimated in the region close to the centre, where both the superfluid and normal components are present, as being 
$\frac{T}{T_c}$ times the total density in this region and we have denoted it by $n_{normal}$ in Eq.~\ref{ekin}. For the Scissor mode we have the full number density denoted by $n$ in the above formulas.

The validity of hydrodynamics imposes a condition on how big $\alpha_x$ can
become, the resulting maximum value, $\alpha_x^{max}$  was estimated in
Eq.~\ref{eq:hydro_condition}.  The quantities $E_{\kinetic}$,
$\dot{E}_{\kinetic}$ which appear in Table \ref{ptrap2} are obtained from
Eq.~\ref{eq:dissipation}, Eq.~\ref{ekin} by setting $\alpha_x=\alpha_x^{max}$. 

A convenient quantity with which to compare $\alpha_x^{\rm{max}}$ is the ratio
of the speed of sound at the centre $c_s=\sqrt{\frac{2\mu}{3m}}$ to a measure
of the trap size $z_{\rm{trap}}$. For comparison, let us note that for
$\omega_z=2 \pi \times 10^4$ rads/s we obtain
$\frac{c_s}{z_{\rm{trap}}}=\frac{\omega_z}{\sqrt{3}}=3.63\times 10^{-11}$ eV.

The (amplitude) damping time $\tau_0$, which appears in Table \ref{ptrap2}, is defined as
\begin{equation}
\label{taudef}
\tau_0 = {2E/\dot{E}_{kinetic}}
\end{equation}

As mentioned above, the table considers the effects of varying the temperature while keeping $\mu, \omega_z, \omega_x,\omega_y$ fixed. 
For fixed $T/ \mu$ one can also consider what happens as the angular frequencies are varied. In the highly anisotropic situations $\omega_z\gg \omega_x, \omega_y$,
one finds that the  total energy $E_{\kinetic}$ for the Elliptic mode approximately scales like
\begin{equation}
\label{ea}
E_{\kinetic}(a) \sim \mu \frac{\mu}{\omega_x}\frac{\mu}{\omega_y}\left(\frac{\mu}{\omega_z}\right)^3
\end{equation}
 and the damping time $\tau_0$ for the Elliptic mode approximately scales like
\begin{equation}
\label{ta}
\tau_0(a)\sim  \frac{\mu}{\omega_z^2}\,.
\end{equation}
Similarly for the Scissor mode we get 

\begin{equation}
\label{eb}
E_{\kinetic}(b) \sim \mu \frac{\mu}{\omega_y}\frac{\mu}{\omega_z}\left(\frac{\mu}{\omega_x}\right)^3,
\end{equation}
\begin{equation}
\label{tb}
\tau_0(b) \sim \frac{\mu }{\omega_x^2}   \,.
\end{equation}
These scalings are obtained by noting that $\alpha_x^{\max}\sim \mu $  for fixed $T/\mu$, and also that the trap potential is unchanged under a rescaling $\omega_z \rightarrow \lambda~ \omega_z, z \rightarrow z/\lambda$ and similarly for $x,y$. We have also assumed that $\omega_z\gg \omega_x, \omega_y$.  Some of these scalings are summarized in the caption below Table \ref{ptrap2}. For example, the scalings of the scissor mode, can be derived as follows: $ E \sim \int dx dy dz [m n v^2] \sim  L_x L_y L_z [m n \alpha^2 L_x^2] \sim \frac{\mu^6} {\omega_x^3 \omega_y \omega_z} $, where we have assumed that at the center of the trap $\mu>0$ and $L_i=\sqrt{2\mu/(m\omega_i^2)}$.) In a similar manner, one can derive the approximate scalings for energy dissipation rates:  $\dot{E} \sim \frac{\mu^5}{\omega_x \omega_y \omega_z}$ for both the modes (assuming $\eta$ scales the same way as $n$ ie. $\sim ( m \mu)^{3 \over 2}$.

The approximate value of $T, \mu, \omega_z$ we consider here are of the same
order as those considered in~\cite{Cao58} where the viscosity of a  unitary
Fermi gas was  measured, using a radial breathing mode. The Scissor mode has been considered in the
literature before.  The damping rate has been measured for cold atoms system in
this mode in superfluid bosonic (see Ref.~\cite{PhysRevLett.84.2056} and Refs. therein) 
and in fermionic systems~\cite{PhysRevLett.99.150403}.  In particular
\cite{PhysRevLett.99.150403} carries out these measurements in the unitary
Fermi gas. The values for trap parameters we consider are similar to
those considered for example in~\cite{Cao58} and not very different from those
considered in~\cite{PhysRevLett.99.150403}. The maximum angular amplitude of the 
the scissor mode is determined by the velocity amplitude $\alpha_x$
(Eqs.~\ref{scissor},~\ref{eq:vprofile}) which is bounded above by 
$\alpha_x^{\rm{max}}$ in Table \ref{ptrap2}. One can show that the angular 
amplitude (in radians) of the oscillation executed by the deformed cloud in the
scissor mode is given by
\begin{equation}
\begin{split}
\theta = \tan^{-1}\left(\frac{e^{\frac{2\alpha_x }{\omega} }-1}{e^{\frac{2\alpha_x }{\omega}}+1}\right)\;,
\end{split}
\end{equation}
where $\omega=\sqrt{\omega_x^2+\omega_z^2}$. Taking $\alpha_x$ to be the
maximum value $\alpha_x^{\rm{max}}\sim 10^{-10}$ eV and $\omega$ to be
$2\pi\times10^{4}$ rads/s $\equiv4.16\times10^{-11}$ eV, we find
$\theta_{\rm{max}}\sim \tan^{-1}[1]$ $\equiv45^\circ$. For a frequency $10$ times
larger, $\theta_{\rm{max}}\sim \tan^{-1}[0.4]$ $\equiv24^\circ$. It is satisfying
that these amplitudes are larger than those measured
in~\cite{PhysRevLett.99.150403} for the scissor mode and hence the condition
for hydrodynamics (Eq.~\ref{eq:hydro_condition}) does not force the amplitudes
to be so small as to preclude observation using existing techniques. 
 For $\mu=10\mu K$, $\omega_x=\omega_y=2\pi\times
385$ rads$/$s and $\omega_z=2\pi\times 10^4$ rads$/$s, $\tau_0$ ranges from
roughly $0.04$ sec to $0.08$ sec. The damping of the scissor mode has been observed for
slightly different parameters values, $\mu\approx 1\mu K$, $\omega_x=2\pi\times 830$ Hz, $\omega_y=2 \pi \times 415$
Hz and $\omega_z=2\pi \times 22$ Hz in Ref.~\cite{PhysRevLett.99.150403} where
the damping time scales measured are of the order of milliseconds.\\


{\it Summary: }

Now we come to the punch line of this section. The effects of anisotropy can
cause a fractional change in components of the viscosity tensor, potentially
lowering some of them.  This effect is expected to go like, $\delta
\eta/\eta\sim \kappa_{{\rm{LDA}}}^2$, as mentioned in
Eq.~\ref{eq:anisotropy_corrections_rough}. We see from Table \ref{ptrap1} that, for
fixed $\omega_z/\mu$, $\kappa_{{\rm{LDA}}}$ increases as $T$ decreases (i.e. $T/\mu$
decreases), with the maximum value, within the  range of allowed temperatures,
being of order $\kappa_{{\rm{LDA}}}\sim 10\%$. This would lead, one expects, to  a
fractional change in components of the viscosity of order $\delta \eta /\eta
\sim (\rm{few}) \times 1 \%$, which is quite small. However note that  increasing
$\omega_z$ will increase $\kappa_{{\rm{LDA}}}$ with a linear dependence
$\kappa_{{\rm{LDA}}}\sim \omega_z/\mu$ as noted in Eq.~\ref{condac} and also in the
first row of Table \ref{ptrap1}.  In turn this should lead to a quadratic
fractional change in  $\delta \eta /\eta\sim ({\omega_z\over \mu})^2$ .  We can
carry out this change while keeping $\omega_x, \omega_y$ fixed thereby
increasing the anisotropy.  Note that this change of $\omega_z$ will decrease
the total energy of this mode $E_{\kinetic}(b)\sim 1/\omega_z$, Eq.\ref{eb},
but it does not change $\tau_0$ significantly, since  $\tau_0$ depends to a
good approximation on  $\omega_x$ and not $\omega_z$ as seen from Eq.~\ref{tb}.
Also note that changing $\omega_z$ while keeping $T/\mu$ fixed will not change
$l$ and thus the localized nature of the region from which the damping arise.
In fact it will make it easier to meet the condition Eq.~\ref{conddd}. 

Also it is worth commenting that  it is easy to see from Eq.~\ref{condac},
Eq.~\ref{eb} and Eq.~\ref{tb}      that  if one want to keep  $\tau_0$ and
$E_{\kinetic}$ for the scissor mode  both fixed and increase
$\kappa_{{\rm{LDA}}} \rightarrow \lambda~ \kappa_{{\rm{LDA}}}$ one could do
this (while keeping $\omega_x=\omega_y$) by scaling 
\begin{equation}
\label{scal}
\omega_x \rightarrow \lambda ^{\frac{1}{6}}~ \omega_x  , ~\omega_y \rightarrow \lambda ^{\frac{1}{6}}~ \omega_y , ~ \omega_z \rightarrow \lambda ^{\frac{4}{3}}~ \omega_z,~ \mu \rightarrow \lambda ^{\frac{1}{3}} ~\mu,~T\rightarrow \lambda ^{\frac{1}{3}} ~T.
\end{equation}
This keeps $\frac{T}{\mu}$,  $\tau_0$ and
$E_{\kinetic}$ fixed, increases the overall magnitude of $\mu$, 
increases $\omega_z$ and also $\omega_x,\omega_y$. 

The discussion of the previous two paragraphs  suggests that  one can quite plausibly keep the damping time scale  and the total energy  in the experimentally accessible range, while gradually increasing $\omega_z$ making $\kappa_{{\rm{LDA}}}\sim \mathcal{O}(1)$ and the effects of anisotropy significant. 
While some of the theoretical approximations made will break down in this limit it is possible  that the effects of anisotropy would get more pronounced, and potentially even dramatic, driving the spin one components of the viscosity to be much smaller than their values in the isotropic case, and potentially even violating the KSS bound.

We have not discussed the Elliptic mode in as much detail. One reason is that unlike the scissor mode, this mode  has not been experimentally realized in cold atom systems yet.\footnote{One possible way
to set up the elliptic mode is to start with a more circular trap and
exciting a rotational mode by using rotating lasers using a set up similar to
Ref.~\cite{2005Natur.435.1047Z}. If the rotational frequency is small enough,
vortices will not be excited and only the normal fluid will rotate like a
rigid body. On adiabatically deforming the trap one would then get the elliptic
mode because during adiabatic deformations, hydrodynamics is satisfied at each 
time and we expect that the normal fluid will go smoothly from circular rotation to 
the elliptic mode. } Also  we see from Table \ref{ptrap2} that the damping time $\tau_0$ in this case is about two orders of magnitude smaller, and this too might be an issue of some experimental concern. 
It may of course turn out that this mode is  experimentally accessible. It will then be certainly interesting to explore its properties, especially since this mode in a very direct way measures the resistance to shear in the resulting fluid flow. 

Finally we note that all the five conditions which were listed at the end of
Sec.~\ref{gravity} for observing the suppression of viscosity can be met in the
system being analyzed here.  Conditions $1$ and $2$ are met by the two modes
discussed above  in the unitary Fermi gas.  We have ensured that $l <1$ (Table
\ref{ptrap1}) so that the contribution arises from a localized region where the
potential is approximately linear, meeting condition $4$. As argued above, for
the scissors mode the anisotropy can be made large enough while staying within
the fluid mechanics approximation ($\alpha_x<\alpha_x^{max}$) thereby meeting
conditions $3$ and $5$. The resulting values for the total energy and the damping
time we find lie  within the experimentally accessible range.

To summarize, we have seen in this section that for  experimentally reasonable
values of parameters  one can   increase the anisotropy of the trapping
potential and probe the   viscosity tensor by measuring the energy loss and
related damping time in the scissor mode. As the anisotropy is increased, its
effects  could well become quite significant driving some components of the
viscosity (spin $1$ in our notation)  to become  very small, and potentially
making them even smaller than the KSS bound. 

\subsection{ Discussion on $\kappa_{{\rm{LDA}}}$}
\label{ kappalda}
In this subsection, we present a detailed discussion on $\kappa_{{\rm{LDA}}}$ given in the last column of Table.~\ref{ptrap1}. The results discussed so far assume LDA is valid. LDA rests on the assumption
that the trap potential varies slowly on the scale of the local Fermi
wavelength $k_{F}^{-1}({\bfr}) = \left(3 \pi^2 n({\bfr})\right)^{\frac{1}{3}}$
ie. at any local point ${\bfr}$ along the length of the trap, the following
condition holds true -
\begin{equation}
\begin{split}
\nonumber \left|{\nabla_{{\bfr}}(\mu(\bfr))} {1 \over k_{F}({\bfr})}\right|_{{\bfr}} 
\ll \mu(\bfr)
 \label{tfgen}
\end{split}
\end{equation}

Since we desire $\omega_x, \omega_y\ll \omega_z$, the gradient is strongest in
the $z$ direction and hence taking $x,\;y=0$ and moving along the harmonic trap
in the $z$ direction, $\frac{d (\mu(z))}{dz} = -m \omega_z^{2}{{z}}$, we note
that LDA violations will be
significant if
\begin{equation}
\begin{split}
 &m\omega_z^2 z\frac{1}{(3 \pi^2 n(z))^{1\over3}}\sim \mu(z)\;.
 \label{tf2}
\end{split} 
\end{equation}

For any trap geometry at the outer edges of the trap when the density becomes
small enough, LDA will be violated ($\mu(z)<0$ for $z>z_{\rm{trap}}$).
These regions typically do not contribute significantly to the trap energy
loss. But focusing on the region near $z_0$, LDA is a good approximation if 
\begin{equation}
\begin{split}
\kappa_{\rm{LDA}} = \frac{\sqrt{m \over 2} \omega_z^{2} z_0}
 {[\calF(\frac{T}{\mu(z_0)})]^{1/3}[\mu(z_0)]^{\frac{3}{2}}} \ll
 1\;,
 \label{tf0}
\end{split} 
\end{equation}
Approximating $ \calF(\frac{T}{\mu(z_0)})]^{1/3}\approx \frac{1}{\sqrt{\xi}}$
[Since $\calF(0)=1/\xi^{3/2}$, and the deviations from $\calF(0)$ are small for
$T/\mu\lesssim 1$], we find 
\begin{equation}
\begin{split}
 \kappa_{\rm{LDA}} =\frac{\sqrt{m \over 2} \omega_z^{2} z_0}
 {[\mu(z_0)]^{\frac{3}{2}}}\sqrt{\xi} \ll
 1\;,
 \label{tf}
\end{split} 
\end{equation}
Since $z_0$ scales as $1/\omega_z$ for fixed $\mu$ and $T$, LDA will be violated at $z_0$ if $\omega_z$ is
large enough. From Table~\ref{ptrap1} one can see that for $\mu=10\mu$K and
$T=T_c/2$, $\kappa_{\rm{LDA}}>1$ for $\omega_z> 2 \pi \times 77000$ rads/s.
Alternatively, taking $\omega_z=2\pi\times10^4$ rads/s and $T=T_c/2$, $\kappa_{\rm{LDA}}$ can become
larger than $1$ if $\mu < 1.3 ~\mu$K.

For $T\rightarrow0$ the corrections to LDA have been previously studied 
in Refs.~\cite{Son:2005rv,Forbes:2012yp}. One can write
\begin{equation}
n(\bfr) = n_{\rm{LDA}}\bigl(1
  -\frac{c_\chi}{64}
   \frac{({\bf{\nabla}} \phi(\bfr))^2+4(\mu-\phi(\bfr)){\bf{\nabla}}^2\phi(\bfr)}
   {m(\mu-\phi(\bfr))^3} +{\calO}{({\bf{\nabla}}^3\phi(\bfr))} \bigr)\;,
\label{eq:LDAviolation0}
\end{equation}
where $c_\chi$ is related to the response of the density to a periodic
fluctuation in the potential. The low energy constant $c_\chi$ has not been
calculated using {\it{ab-initio}} techniques so far. In all 
model calculations $c_\chi\sim 1$, including in a
sophisticated analysis using SLDA (Ref.~\cite{Forbes:2012yp}).

For finite $T$ for an isothermal system, the deviations from LDA are not
related to the density response but for $T\lesssim (\mu-\phi(\bfr))$ we can
write corrections to LDA in analogy with Eq.~\ref{eq:LDAviolation0} 
\begin{equation}
n(\bfr) = n_{\rm{LDA}}\bigl(1
  -\frac{c_1}{64}
   \frac{({\bf{\nabla}} \phi(\bfr))^2}
   {m(\mu-\phi(\bfr))^3}
  -\frac{c_2}{16}
   \frac{{\bf{\nabla}}^2\phi(\bfr)}
   {m(\mu-\phi(\bfr))^2}  + {\cal{O}}((\nabla V)^3) \bigr)\;,
\label{eq:LDAviolation}
\end{equation}
where $c_{1,\;2}$ are functions of $(T/\mu)$ and tend to $1$ as
$T/\mu\rightarrow 0$. In particular, for the
interesting region the term proportional to $c_1$ is dominant (the exception
is near the center of the trap). Therefore, the corrections to LDA near $z_0$
can be written as
\begin{equation}
n(z) = n_{\rm{LDA}}\bigl(1
  -\frac{c_1}{64}\frac{2}{\xi}
  \kappa^2_{\rm{LDA}} +\cdot\cdot\bigr)\;,
\label{eq:LDAviolationvskappa}
\end{equation}
where we have used the low temperature expression
\begin{equation}
m\mu(\bfr) = \frac{\xi}{2} k_F^2(\bfr)\;,
\end{equation}
to write the correction in terms of $\kappa_{\rm{LDA}}$.

In the absence of further information about $c_1$ at finite $T$ it is difficult
to make precise statements about the relevance of LDA corrections for the traps
with large values of $\omega_z$ that we show in the next Section are needed to
make the shear viscosity tensor locally anisotropic. Therefore, we simply use
$\kappa_{\rm{LDA}}\gtrsim 1$ as a marker for significant LDA violation.
However, it is important to keep in mind that if $c_1(\frac{T}{\mu(z_0)})\sim
c_1(0.54) \sim1$ (since $\frac{T}{\mu(z_0)}\sim 0.54$ for the cases we
consider), then the pre-factor of $1/(32\xi)$ implies that the corrections to
LDA can be small even for $\kappa_{\rm{LDA}}\approx 1$.

\section{Local anisotropy}
\label{local}

Hydrodynamics is an effective theory: The conserved currents are written as a
series of terms ordered by the number of  derivatives acting on the local fluid
velocity. The lowest order terms are simply given by the Galilean (for
non-relativistic systems) or Lorentz (for relativistic systems) transforms of
the local thermodynamic properties like the density and the pressure, from the
local rest frame of the fluid to the laboratory frame. The first order terms
are given by the local gradients of the velocity $(\partial_i u_j +\partial_j
u_i)/2$ multiplied by proportionality constants given by the transport
coefficients --- for example viscosities --- of the system. We will not
consider higher derivative terms in this paper, instead restricting ourselves
to situations (see Eq.~\ref{eq:hydro_condition1}) where the first order
correction is smaller than the lowest order terms.
 
In the presence of external fields, the law of conservation of energy features
a source term proportional to the driving force, $\nabla \phi(\bfr)$. If
$\nabla \phi(\bfr)$ is ``small'' (which we shall define in a moment), its effect on the
thermodynamics and transport can be neglected, and hydrodynamics describes a
locally isotropic fluid (with isotropic thermodynamic functions and isotropic transport
coefficients)~\footnote{This assumes that microscopically the fluid is
isotropic. For example it is not a crystal~\cite{LandauPhysical} or a fluid phase with an
anisotropic order parameter.} moving in a space dependent potential. The key
realization therefore is that to observe an anisotropy in thermal or transport
properties it is not sufficient for $\omega_x, \omega_y\ll \omega_z$.
Corrections to isotropy will start becoming significant as we increase
$\omega_z$, if $\omega_z$ starts becoming comparable to some microscopic scale
of the system. 

The criterion for the thermodynamic quantities to exhibit the effect of $\nabla
\phi(\bfr)$ is clear from the previous section. If the potential varies on
length scales comparable to the inter-particle separation --- the Thomas-Fermi
approximation, or LDA breaks down --- the pressure of the fluid in the
direction of the gradient will be different from the pressure in the
perpendicular directions. In this case, clearly the transport coefficients will
also be anisotropic. To explore an analogous system to the one described in
Sec.~\ref{gravity}, this argument prompts us to consider $\omega_z$ large
enough that LDA is broken (see Table~\ref{ptrap1}). For such systems, the
estimates for the density Fig.~\ref{num} and viscosities Fig.~\ref{localvis}
using LDA will be only rough guiding values, but if the analogy with the system
in Sec.~\ref{gravity} holds true, the viscosity values relevant for the modes
described in Sec.~\ref{vprofile} will be lower than the LDA values, and could
be lower than $1/(4\pi)$ in suitable quantum units.

To estimate the order of the correction to the shear viscosity due to potential
gradients we note that the first order correction to transport due to $\nabla
\phi(\bfr)$ simply appear as the source term, and hence assuming that the next 
order corrections will be analytic in $\nabla \phi(\bfr)$, we expect 
\begin{equation}
\eta_{ijkl} = 
\eta \frac{1}{2}
[(\delta_{ik}\delta_{jl}+ \delta_{il}\delta_{jk}-
\frac{2}{3}\delta_{ij}\delta_{kl})
+
\bigl(\frac{\lambda^2(\nabla \phi(\bfr))(\nabla
\phi(\bfr))}{[\mu(\bfr)]^2}\bigr)\sum_{\alpha=0}^{4}
c_{(\alpha)}M_{\alpha\,ijkl}]+\calO(\nabla^2\phi, (\nabla\phi)^4)\;,
~\label{eq:anisotropy_corrections0}
\end{equation}
where $\lambda$ is a microscopic length scale of the system, $c_{(\alpha)}$ are dimensional constants of
order $1$ which depend on the microscopic details of the system, and
$M_i$ are $5$ orthonormal projection operators that arise in a system
with one special direction (for eg. see Ref.~\cite{Goossens2010}). We have
given these projection operators in Appendix.~\ref{microscopic} (Eq.~\ref{eq:projections}). 

$\lambda$ is a length scale that determines transport behavior. In a system
admitting a quasi-particle description we expect $\lambda$ to be of the order
of the mean free path. (We show this explicitly in
Appendix.~\ref{microscopic}.) The other length scale in the system is the
inter-particle separation $1/k_F$. In terms of $k_F$ we can write the
corrections as 
\begin{equation}
\begin{split}
\eta_{ijkl} &\approx 
\eta \frac{1}{2}
[(\delta_{ik}\delta_{jl}+ \delta_{il}\delta_{jk}-
\frac{2}{3}\delta_{ij}\delta_{kl})
+ 
(\lambda k_F)^2\bigl(\frac{(\nabla \phi(\bfr))(\nabla
\phi(\bfr))}{k_F^2[\mu(\bfr)]^2}\bigr)\sum_{\alpha=0}^{4}
c_{(\alpha)}M_{\alpha\,ijkl}]\\
&=
\eta \frac{1}{2}
[(\delta_{ik}\delta_{jl}+ \delta_{il}\delta_{jk}-
\frac{2}{3}\delta_{ij}\delta_{kl})
+
(\lambda k_F)^2\bigl(\kappa^2_{\rm{LDA}}\bigr)\sum_{\alpha=0}^{4}
c_{(\alpha)}M_{\alpha\,ijkl}]\;,
~\label{eq:anisotropy_corrections}
\end{split}
\end{equation}

For weakly interacting quasi-particles, the $\lambda k_F\gg1$. But for a
strongly interacting system in the absence of more information about $\lambda
k_F$ and $c_{(\alpha)}$ it is not possible to make a more concrete statement about
the corrections to viscosity.  We can only state that the corrections are
important if $\kappa_{\rm{LDA}}\sim1$ as we did in Eq.~\ref{eq:anisotropy_corrections_rough}.

As discussed in Sec.~\ref{gravity}, for the theories considered in
Sec.~\ref{gravity}, there is no quasi-particle description. The only relevant
length scale is $1/T$ and the field $\phi$ changes by order $1$ on a length
scale $1/\rho$. Using AdS/CFT it has been shown~\cite{Jain:2015txa} that the
corrections to isotropy go as Eq.~\ref{eq:eta_low}. 

For the unitary Fermi gas there is no known gravitational
dual~\cite{Bekaert:2011cu} and we will need to resort to a rough calculation to
estimate $c_{(\alpha)}$ and $\lambda k_F$. We solve the Boltzmann transport equation
in the relaxation time approximation.  We hope this will give semi-quantitative
results. We leave the challenging calculation of the viscosity for temperatures
in the strongly coupled regime just above the critical temperature in the
presence of a background potential for future work.

As we show in Appendix.~\ref{microscopic}, the corrections to $\eta$ for a
weakly interacting, normal (unpaired) Fermi gas at low temperatures ($T<\mu$) 
are given by (Eq.~\ref{eq:etaoflambda})
\begin{equation}
\begin{split}
\eta_0 &= \eta(0)[1-\frac{31}{84}(\lambda k_F)^2
    \frac{(\E)^2}{k_F^2\mu^2}+\calO((\tau\E)^4)]
    =\eta(0)[1-\frac{31}{84}(\lambda k_F)^2\kappa_{\rm{LDA}}^2+\calO((\tau\E)^4)]\\
\eta_1 &= \eta(0)[1-\frac{13}{28}(\lambda k_F)^2
    \frac{(\E)^2}{k_F^2\mu^2}+\calO((\tau\E)^4)]
    =\eta(0)[1-\frac{13}{28}(\lambda k_F)^2\kappa_{\rm{LDA}}^2+\calO((\tau\E)^4)]\\
\eta_2 &= \eta(0)[1-\frac{11}{28}(\lambda k_F)^2
    \frac{(\E)^2}{k_F^2\mu^2}+\calO((\tau \E)^4)]
    =\eta(0)[1-\frac{11}{28}(\lambda k_F)^2\kappa_{\rm{LDA}}^2+\calO((\tau\E)^4)]\\
\eta_3 &= 0, ~\eta_4 = 0\;,~\label{eq:etaoflambda2}
\end{split}
\end{equation}
where $\tau$ is the effective relaxation time.\\
For the Elliptic mode $\frac{1}{2}(\partial_i u_j + \partial_j
u_i)=\frac{1}{2} \alpha_x(1-\frac{\omega_x^2}{\omega_z^2})= V_{xz}$ which
probes the viscosity contribution to the stress energy tensor 
\begin{equation}
\sigma_{2_{ \alpha \beta}}  = 2 ~\eta_2~(V_{\alpha \gamma} b_{\beta} b_{\gamma} +b_{\alpha}V_{\beta \gamma} b_{\gamma}- 2 b_{\alpha} b_ \beta b_\gamma b_\delta V_{\gamma \delta} )
\;,
\end{equation}
where $b$ is a unit vector along the gradient of the potential.
For the Scissor mode, $\frac{1}{2}(\partial_i u_j + \partial_j
u_i)=\alpha_x = V_{xz}$ which also probes $\eta_2$. ($\eta_2$ is the
coefficient that corresponds to the projection operator $M_2$ in Eq.~\ref{eq:projections}.) 

In both cases (see Appendix.~\ref{microscopic}) , $\eta$ is reduced from its value in the absence of the
potential, $\eta(0)$, for $\frac{\tau^2}{k_F^2}(\E)^2\lesssim 1$. To estimate the
value of $\tau$ near $z=z_0$, we note that for $z\sim z_0$, $T(z_0)\sim
0.54~ \mu(z_0)$. At this $T$, $\eta(0)/n|_{z_0}\sim 1$. 

Using the relaxation time approximation and
thermodynamic expressions for a weakly interacting Fermi gas to estimate 
$\lambda$ near $z_0$, we obtain (Eq.~\ref{eq:eta0degenerate}) 
\begin{equation}
\begin{split}
{\eta(0)}(z_0) &= \frac{(2m\mu(z_0))^\frac{5}{2}\tau(z_0)}{15\pi^2 m}\\
&=\frac{2}{5}n(z_0)\mu(z_0)\tau(z_0)\;.
\end{split}
\end{equation}
Therefore near $z_0$, $\tau(z_0)\sim
\frac{5}{2\mu(z_0)}\frac{\eta(0)}{n}|_{z_0}$, or, 
\begin{equation}
\begin{split}
\lambda(z_0) &= v_F(z_0) \tau(z_0)\\
&\sim \frac{k_F(z_0)}{m} \frac{5}{2\mu(z_0)}\frac{\eta(0)}{n}|_{z_0}\\
&=\frac{5}{4k_F(z_0)}\frac{\eta(0)}{n}|_{z_0}\;.
\end{split}
\end{equation}
(We have just kept the pre-factors of the order of $1$ to serve as mnemonics of
the derivation of $\lambda$. They have no quantitative significance.)

Therefore, (since $\frac{\eta(0)}{n}|_{z_0} \sim 1$ from $\frac{\eta}{n}$ data)
\begin{equation}
\begin{split}
 \lambda(z_0)k_{F}(z_0)  =\frac{5}{4}\frac{\eta(0)}{n}|_{z_0}\sim 1\;. 
 \end{split}
\end{equation}

The fact that $k_{F}(z_0) \lambda(z_0)\sim 1$ means that the Boltzmann
transport calculation shown in Appendix.~\ref{microscopic} is not
quantitatively trustworthy near $z_0$. But we hope that two the main qualitative 
consequences of Eq.~\ref{eq:etaoflambda2} survive a more controlled calculation.
\begin{enumerate}
\item{First, the coefficient of $\kappa_{\rm{LDA}}^2$ in
Eq.~\ref{eq:etaoflambda2} is of the order of $1$.}
\item{Second, the sign of the correction term is negative}
\end{enumerate}
If true, this would imply that the shear viscosity component $\eta_{xzxz}$
measured using the Elliptic mode or the Scissor mode will reduced by order $1$
from its value in isotropic traps, if 
$\omega_z\gtrsim  2 \pi \times 77000$ rads/s (Table.~\ref{ptrap1}).

One might be concerned that for $\omega_z\sim 2 \pi \times 77000$ rads/s, our
conclusions in the previous section about $\delta z/z_0$ will be violated
because of the violation of LDA. In the absence of more concrete information on
these coefficients we can not assure this will not happen. We simply note that
if the coefficient $c_1$ in Eq.~\ref{eq:LDAviolationvskappa} is of the order of
$1$ (which it is at $T\ll\mu$, but may be larger for $T\sim0.54~\mu(z_0)$) then there
is a regime where the corrections to the thermodynamics due to LDA is small,
but the reduction in transport coefficients is substantial. 

\section{Conclusions}
\label{res}

We present a concrete realization of a system of ultra-cold Fermi gases at
unitarity, in an anisotropic trap, which may show significant
reduction in the viscosity compared to its value in isotropic traps. Given that
the value of the isotropic viscosity has been measured to be few times the KSS
bound in this system, it presents a candidate setup  to observe a shear viscosity
 smaller than the KSS bound when it is subjected to an anisotropic driving force.

The anisotropic force is obtained by placing the system in an anisotropic trap.
The trapping potential is harmonic, Eq.~\ref{eq:harmonic_potential}, and characterized by three angular frequencies, $\omega_x,\omega_y,\omega_z$. 
We consider an anisotropic situation where $\omega_z\gg \omega_x, \omega_y$, so that the trapping potential is much stronger in the $z$ direction.
For simplicity, we also take $\omega_x=\omega_y$ so that the system preserves rotational invariance in the $x-y$ plane.
For some of the discussion below we can neglect the effects of the trapping potential in the $x,y$ directions characterized by  $\omega_x,\omega_y$. 

We work in conventions where $k_B=\hbar=1$. 
There are three energy scales $T, \mu, \omega_z$  and two dimensionless ratios $T/\mu$ and $\omega_z/\mu$ which then characterize the system. 
The Li$_6$ atoms have a mass $m$, using this parameter,  any of the  energy scales  can be converted  to a length scale, $L={1\over \sqrt{2 m E}}$. 

Based on the behavior seen quite generically in gravity systems we identify
five criterion (Sec.~\ref{cond}) which when met could plausibly lead to a
decrease in the value of some components of the  viscosity tensor (the spin one
components).  These are summarized towards the end of  Sec.~\ref{gravity} .  On
studying the superfluid equations we identify two modes which are sensitive to
these components of the viscosity tensor.  One of these  is the scissor mode
which has already been studied experimentally in some detail.  By taking
reasonable values for  the parameters- $T$, $\mu$, $\omega_z, \omega_x,
\omega_y$, which are in the experimentally accessible range, Ref.~\cite{Cao58},
we find that all the five  criteria can be met.  Furthermore, we find that the
resulting   energy and damping rate of this energy, from which the viscosity
can be extracted, lie within the range of values which are measured by
experiments currently being done on cold atom systems,  in particular on Li$_6$
unitary Fermi gas systems, Ref.~\cite{PhysRevLett.99.150403}.  For example, for
$\mu=10\mu$K, $\omega_z \sim 2 \pi \times 77000$ rads/s, and $T= \frac{T_c}{2}$
($T_c=0.4\mu$) we find that the anisotropy, as measured by the parameter
$\kappa_{LDA} $ , Eq.~\ref{defkappalda}, is of order unity and therefore
significant.  At these extreme values of anisotropy our theoretical
calculation, strictly speaking, do not apply, but a reasonable extrapolation
suggests that  the maximum total energy is of the order of $10^{-17}$ joules which corresponds to the angular
amplitude of the scissor mode of about $24^{\circ}$ which is within the experimental range of
\cite{PhysRevLett.99.150403}. The damping time $\tau_{0}$ is of the order of $10^{-2} $ seconds, which is roughly ten times longer than the observed
amplitude damping time that has been accurately measured in the experiments on
ultracold Fermi gases \cite{PhysRevLett.99.150403}.
\\
While the system is certainly close to being two-dimensional when $\kappa_{LDA} \sim1$ and $z_{trap}  \sim 5.4~ k_F^{-1}$ (this corresponds to $\mu/\omega_z \sim 2.7$) is on the small side,  the effect of small viscosity can already set in when $\kappa_{LDA}$ is somewhat smaller than unity. 
 We illustrate this with concrete quantitative examples below. \\
 For concreteness, let us consider traps where we fix $T/T_c=1/2$ ($T_c=0.4\mu$,
where $\mu$ is the chemical potential at the center of the trap) and change
$\omega_z$. Further, for concreteness, we set the overall scale by
$\mu=10\mu$K. Considering first a representative trap geometry where the shear viscosity
tensor is locally isotropic to a large accuracy, we take $\omega_z=0.048\mu$
(corresponding to $\omega_z=2\pi\times 10^4$Hz which is typical), for which
$\kappa_{\LDA}=0.13$.  The fractional reduction in the shear viscosity for this
value of $\omega_z$, taking $c_2$ to be its Boltzmann transport value $11/28$ is
\begin{equation}
\begin{split}
\frac{\Delta\eta}{\eta}\approx -\frac{11}{28} (\kappa_{\LDA})^2 = -0.7\%\;,
\end{split}
\end{equation}
which is a small reduction in the shear viscosity and may not be even
measurable above measurement errors. At the other extreme we considered, $\omega_z= {\mu \over 2.7 }$ (corresponding to
$\omega_z=2 \pi \times 77.16$~kHz), for which $\kappa_{\LDA}=1$ and the fractional reduction is   
\begin{equation}
\begin{split}
\frac{\Delta\eta}{\eta}\approx -\frac{11}{28} (\kappa_{\LDA})^2 = -39\%\;,
\end{split}
\end{equation}
 which is very large. However,
in this extreme limit ($\omega_z={ \mu \over 2.7 }$) only the lowest $2-3$ Landau levels are occupied and the
dynamics may be approximately two dimensional. Now consider an intermediate value, say $\omega_z=0.9T = 0.18 \mu$ for which $\kappa_{LDA}= 0.48 <1$. This   gives a correction  
\begin{equation}
\begin{split}
\frac{\Delta\eta}{\eta}\approx -9\% 
\end{split}
\end{equation}
which --- while not large --- is still substantial. More generally,  the criterion for confinement in the $z$ direction is 
\begin{equation}
\begin{split}
\omega_z\gtrsim {\rm{max}}(\Delta, T)\;,
\end{split}
\end{equation}
since both $T$ and pairing allow for excitations between the harmonic oscillator levels. At these extreme values,  where the inequality above is met, our approximations do break down, (shell effects become important as $\omega_z\gtrsim T$, which is another way of saying that confinement in the $z$ direction becomes strong). For $\omega_z={ \mu \over 2.7 }$, $\omega_z=1.85~ T$ and indeed confinement in the $z$ direction is too strong. But, as illustrated by the cases above,  by taking $\omega_z$ a factor of $2$ or $3$ smaller ( say  $\omega_z=0.9~ T$  that was chosen above for illustration\footnote{The deviations
from LDA due to shell effects for unpaired fermions was calculated in Ref.~\cite{PhysRevA.67.053601}. A naive
application of the results of Ref.~\cite{PhysRevA.67.053601} suggests that for our trap with $\omega_z =0.18\mu$, the corrections to the 
number density is about $15\%$ at $T=4 T_c/5$ near the region relevant for our purposes. Note however, that pairing
suppresses LDA violations (\cite{Forbes:2012yp,Bulgac:2010dg}) and we expect the corrections to be much smaller in the relevant
region.}) than the extreme limit, one can measure the tendency of the spin one component of the viscosity to decrease from its lowest value observed in ultra-cold Fermi gases. In an optimistic scenario where $c_2$ is larger in magnitude than the approximate value of $11/28$ in the Boltzmann transport approximation, the reduction will be even more substantial. Let us also point out that comparing with Ref.\cite{PhysRevLett.108.070404} the typical values of $\omega_z/E_F$ in the paper is about $80$ and the value of $\omega_z/T$ is $120$. In that case, the trap is truly 2 dimensional as opposed to when $\omega_z/T\sim 0.9$.\\
Thus,  for smaller values of anisotropy, the
theoretical estimates  are more reliable and suggest that the different
viscosity tensor components should have a fractional difference given in terms
of $\kappa_{LDA}$ by Eq.~\ref{eq:etaoflambda2}. This tendency of the
viscosity to decrease should already be measurable at more moderate values of
the anisotropy. 

Our proposal is the first proposal to measure parametrically suppressed
anisotropic viscosity components in ultra-cold Fermi gases. Our proposal is
different from the discussion of anisotropic hydrodynamics in
Ref.~\cite{Bluhm:2015raa} since we are demanding that hydrodynamics be a good
description (in the sense of Eq.~\ref{eq:hydro_condition1}) in the regime which dominantly contributes to the dissipation of
the fluid dynamics modes.

Future theoretical work can improve upon our proposal in several ways. First,
our estimate of the corrections to the shear viscosity components due to the
potential (Eq.~\ref{eq:etaoflambda2}) was based on a relaxation time treatment
of the Boltzmann equation. For strongly interacting fermions, this is not a
good approximation and a more rigorous calculation of the anisotropy
corrections is desirable. This will require calculating transport properties in a strongly coupled theory
without a gravitational dual, in the presence of a background potential: a
formidable challenge. Second, we have focused on the region that dominantly
contributes to the dissipation. In particular we have neglected the contributions from the tail of
the cloud. While this is presumably small, it would be nice to establish this by
solving the Boltzmann transport equations in this dilute regime.

It is also worth noting   that while the cold-atom system proposed here shares
many features with those discussed in Ref.~\cite{Jain:2014vka,Jain:2015txa}, it
also has some differences. First, in equilibrium the stress energy tensor is
not invariant under translations even for a linear potential. Rather the
density decreases with increasing $z$, but the driving force is proportional to
the gradient of the potential $\phi(\bfr)$ (see Eq.~\ref{eq:hydrodynamics}) as
in Ref.~\cite{Jain:2014vka,Jain:2015txa}. Second, in addition to
energy-momentum, the cold-atom system features another conserved quantity: the
particle number.  Consequently the system is locally characterized by two
thermodynamic variables $T$ and $\mu$ rather than just $T$. It would also be
interesting to  further study the behavior of viscosity in gravitational
systems which correspond to  anisotropy driven strongly coupled systems with
a finite chemical potential. The examples in
Ref.~\cite{Jain:2014vka,Jain:2015txa} did not have a finite chemical potential,
for some discussion of anisotropic gravity systems with a chemical potential
see Ref.~\cite{Ge:2014aza,Chakraborty:2017msh}. As a first step, we have analyzed a weakly
coupled system with a linear varying  potential in Appendix.~\ref{microscopic} and find that the
viscosity does become anisotropic in this case.\\

However, there is no reason to wait for these theoretical advances.  The
central point of this paper  is that there is already enough motivation, based
on the behavior quite generically seen in gravitational systems, to suggest
that some components of the viscosity tensor in anisotropic strongly coupled
systems  might well become small, making $\eta/s$ for these components
potentially even smaller than the KSS bound,  $1/ 4\pi$. Such a decrease in the
viscosity might well  happen in cold atom systems, for example the unitary
fermi gas, which are experimentally well studied. As argued above,  the range
of values involved for temperature, chemical potential and angular frequencies
are well within the experimental regime for such a system, and the  scissor
mode which is sensitive to the relevant  components of the viscosity has
already been realised  experimentally in them. Further, the resulting values
for the energy and the damping time from which the viscosity can be extracted
lie in the experimentally accessible range which has already been achieved. 
  
 We hope our experimental colleagues in the cold atoms community will take note
of these results, and implore them to carry out a study of viscosity in
anisotropic traps.
  
\section{Acknowledgments}
We thank D. D. Ofengeim and in particular D. G. Yakovlev for sharing their
notes on the calculation of the various components of viscosities in the
presence of the magnetic field. We especially thank M. Randeria for sharing
his valuable comments and insights.  We also acknowledge conversations with K.
Damle, S. Gupta,  S. Jain,  N. Kundu, G.  Mandal, S. Minwalla, T. Sch{\"a}fer, R. SenSarma and N. Trivedi.
SPT acknowledges support from the J. C. Bose fellowship DST, Government of
India.  We acknowledge support from the DAE, Government of India. Most of all
we thank the people of India for supporting our research.

\appendix
\section {More details on the results of shear viscosity from gravity}
\label{grdetails}
In Ref.~\cite{Jain:2015txa}, several anisotropic theories in $3+1$ dimensional
space-time (the boundary with coordinates $(t,x,y,z)$), which are dual to a
gravitational theory living in $4+1$ dimensional space-time (the bulk with an
additional coordinate $u$) were studied. Isotropy was broken by considering
states where some of the fields have a background value that depended on some
of the spatial coordinates $x\;,y\;,z$, explicitly breaking rotational
symmetry between them.  

All the examples studied in Ref.~\cite{Jain:2015txa} share the common feature
that the force responsible for breaking isotropy in the boundary theory is
translation invariant as we shall explain via an example below. 

Ref.~\cite{Jain:2015txa} built on the results of Ref.~\cite{Jain:2014vka},
which studied a simple system consisting of a  linearly varying dilaton. The  dilaton field $\phi$ 
couples to the graviton in the bulk via the Lagrangian
\begin{equation}
S = \frac{1}{16\pi G} \int d^5 x \sqrt{g}~ [R+12\Lambda -
\frac{1}{2}\partial_\mu \phi\partial^\mu\phi]\;,~\label{eq:5dlag}
\end{equation}
where $G$ is Newton's constant in $5$ dimensions and $\Lambda$ is a cosmological
constant. The boundary theory in the absence of anisotropy is a $3+1$ dimensional
conformal field theory. 

In this system we can clarify what we mean by saying that the driving force is constant. The dilaton field in the background solution here has the profile
\begin{equation}
\label{anisoparama}
\phi(t,x,y,z)=\rho z\;.
\end{equation}
Clearly this choice of the background singles out the  $z$ direction, breaking
isotropy. In the presence of the dilaton the conservation equations for the stress tensor get modified to be,
\begin{equation}
\partial_\mu T^{\mu\nu}=\langle O \rangle \partial^\nu \phi~\label{eq:hydrodynamicsa}\;,
\end{equation}
where $O$ is the operator dual to the field $\phi$. 
The right hand side arises because the varying dilaton results in a driving force on the system.
We see that a linear profile results in a constant value for $\partial^\nu \phi$ and thus a constant driving force. 

Let us also mention that in this example, on the gravity side the linearly varying dilaton gives rise to a translationally invariant  stress tensor and thus a black brane solution which preserves translational 
invariance. This corresponds to the fact that in the field theory the equilibrium stress tensor features only derivatives of $\phi$ and is thus
space-time invariant.



We shall see that the cold-atom system we consider will not be invariant
under translations in equilibrium. However the equations of hydrodynamics (Eq.~\ref{g_cons}) in the presence of a driving force associated
with a space varying potential look similar to Eq.~\ref{eq:hydrodynamicsa},
where the operator $O$ in the cold-atom system corresponds to the density, and
the driving force is proportional to the gradient of the potential
$\phi(\bfr)$.

The example considered in Ref.~\cite{Jain:2014vka} also shares the
property that an $SO(2,1)$ residual Lorentz symmetry survives, at zero temperature,
after breaking isotropy. This residual Lorentz symmetry corresponds to the $t,\;x,\;y$
directions in the boundary theory.
Fluid mechanics corresponds to the dynamics of the
Goldstone modes associated with the boost symmetries of this residual Lorentz
group, which are broken at finite temperature.

In a general system the viscosity $\eta$ is a fourth order tensor under
rotations relating the deviation of the stress-energy tensor from its
equilibrium value, to the velocity gradient. If the local fluid velocity is
$\bfv=(v_x,v_y,v_z)$, we have
\begin{equation}
\delta T^{ij} = \eta^{ijkl} \frac{1}{2}(\partial_k v_l+\partial_l v_k)\;.
\end{equation}
Since we are only considering the effects of the shear components, 
\begin{equation}
\eta^{ijkl} \delta_{kl} = 0\;.
\end{equation}

\begin{figure}
\begin{center}
\includegraphics[width=0.5\textwidth]{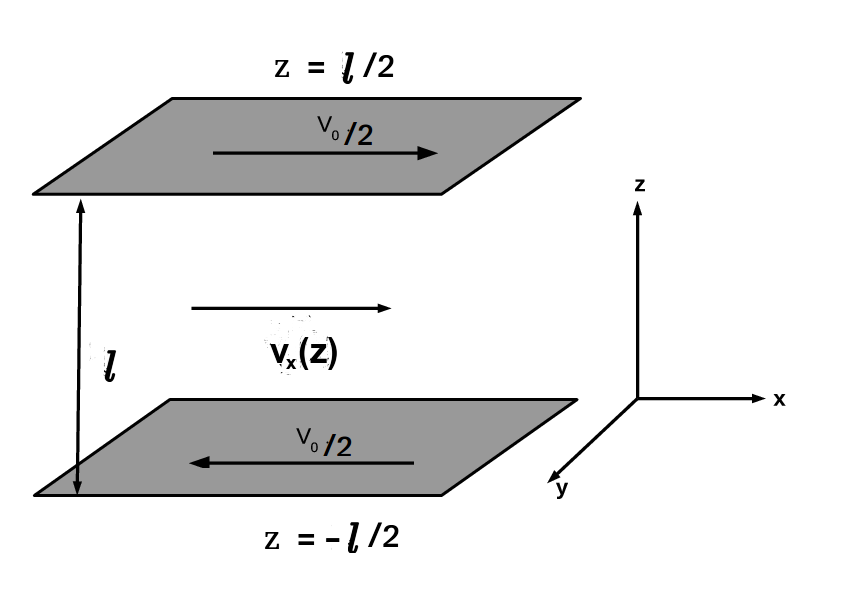}
\caption{Fluid flow between two parallel plates. For $\phi=\rho z$ the driving
force is in the $z$ direction and is proportional to $\rho$. Parametrically
small values of the viscosity (Eq.~\ref{eq:eta_higha}) govern the dynamics for
flows in the $x$ (or $y$) direction with a gradient in the $z$ direction (for
Eg. $v_x=v_0 z$).}
\label{plates}
\end{center}
\end{figure}

In the example in Ref.~\cite{Jain:2014vka}, with dilaton profile given by Eq.~\ref{anisoparama}, the viscosity components that  become small  correspond to  the $\eta^{xzxz}, \eta^{yzyz}$ 
components of the viscosity tensor. In the subsequent discussion we shall use an abbreviated notation,
\begin{equation}
\label{abbota}
\eta^{xzxz}=\eta_{xz}, ~   \eta^{yzyz}=\eta_{yz}.
\end{equation}
In the gravity description these components correspond to perturbations  of the metric
which carry spin $1$ with respect to the surviving $SO(2,1)$  residual  Lorentz symmetry.

A fluid flow configuration where the frictional force (and therefore the resulting dissipation) is governed by a  spin $1$ viscosity component arises  as follows. 
Consider the fluid  enclosed between
(\cite{Jain:2015txa,LandauPhysical}) two parallel plates separated
along $z$ axis by a distance $L$ with the top plate moving with a speed
$v_{0}/2$ along $x$ direction while the lower plate moves with a speed
$v_{0}/2$ along $-x$ direction, see Fig.\ref{plates}. 

The resulting steady state solution of the Navier Stokes equation, even for the anisotropic case,  is remarkably simple,  with 
\begin{equation} 
v_{y} =0,\; v_{z} =0,
\end{equation} 
the  temperature $T$ being  a constant, and $v_{x}$  being a linear function of $z$
\begin{equation} 
v_{x} ={ v_{0} \over L} z,\; z\in(-L/2,L/2)
\end{equation} 
 ( we have chosen coordinates so that $z=0$
lies at the midpoint between the plates).  A constant force per unit area  is exerted by the fluid  on both the upper and lower plates, 
$T^{xz} = \eta_{xz}~ \partial_{z}v_{x}$, in this solution (we
are compactly writing $\eta_{xzxz}$ as $\eta_{xz}$).  This frictional force retards the relative motion of the plates and must be counteracted by an equal and opposite force acting on both plates externally to sustain the steady state solution \footnote{It is interesting to note that this
solution for $\bfv$ is essentially the same as that of an isotropic fluid.}. We also note that for this solution, in the gravity theory under discussion, 
hydrodynamics  is  valid as long as the velocity gradient $v_{0} \over L$ is
small compared to the temperature $T$.

Using results from the gauge-gravity duality~\cite{Maldacena:1999} it was shown
in Ref.~\cite{Jain:2015txa} quite generally that the viscosity component $\eta_{xz}$ behaves
like
\begin{equation}
\label{mresaa}
{\eta_{xz}\over s} = {1\over 4\pi} {{g}_{xx}\over { g}_{zz}}~\Big{|}_{u=u_{h}}, 
\end{equation}
where ${ g}_{xx}|_{u=u_{h}}, {g}_{zz}|_{u=u_{h}}$ refer to the components of
the background metric evaluated at the horizon which we denote by $u_{h}$. `$s$' refers to the entropy density which in the bulk picture corresponds to the area of the event horizon.

In the isotropic case the ratio ${{g}_{xx}\over {g}_{zz}}~\Big{|}_{u=u_{h}}$ is
unity and we see that the KSS result is obtained.  However, in anisotropic
cases this ratio can become very different from unity and in fact much smaller,
leading to the parametric violation of the KSS bound, where the relevant
dimensionless parameter is the ratio of the strength of the anisotropic
interaction and an appropriate microscopic energy scale of the system.

The general  result Eq.~\ref{mresaa}, for  the behavior of the spin $1$ shear viscosity  components $\eta_{xz}=\eta_{yz}\equiv \eta_{\perp}$ was studied in the example of  Ref.~\cite{Jain:2014vka}  for two
cases --- one in the low anisotropy regime and the other in the high anisotropy
regime. 
In this example, there are two scales of interest, $\rho$, which enters in the dilaton profile, Eq.~\ref{anisoparama} and determines the anisotropy, and the temperature $T$ (while this theory 
does 
not have quasi-particles at finite $T$, one can roughly think of the mean free
path as being of the order of $1/T$). Whether the anisotropy is large or small is determined by  the ratio $\rho/T$ which is dimensionless.  Simple results can be obtained in the limit of low
and high anisotropy which correspond to $\rho/T \ll 1$ and $\rho/T\gg 1$ respectively. 

For the spin $1$ component of the shear viscosity $\eta_{xz}=\eta_{yz}\equiv \eta_{\perp}$ the results are as follows:

\begin{enumerate}
\item{Low anisotropy regime ($\rho/T \ll 1$):
\begin{equation}
\frac{\eta_{\bot}}{s}=
\frac{1}{4\pi}-\frac{\rho^2 \log 2}{16 \pi^3 T^2}
+
\frac{(6-\pi^2+54 (\log 2)^2)\rho^4}{2304\pi^5 T^4}
+
{\cal{O}}\bigg[\bigg(\frac{\rho}{T}\bigg)^6\bigg]~\label{eq:eta_lowa}\;.
\end{equation}
We see that a small anisotropy  at order $(\rho/T)^2$ already reduces this component of the viscosity and makes it smaller than the KSS bound.
In the limit of zero anisotropy, we recover the KSS bound
\begin{equation}
\frac{\eta_{\bot}}{s}\rightarrow\frac{1}{4\pi}.
\end{equation}
We also note that the driving force in the conservation equation for the stress
tensor (Eq.~\ref{eq:hydrodynamicsa}) is proportional to $ \nabla \phi\sim \rho$
(Eq.~\ref{anisoparama}) and the analogue of the mean free path is T. Thus the
corrections go like $\frac{(\nabla \phi)^2}{T^2}$.}
\item{High anisotropy regime ($\rho/T \gg 1$):

\begin{equation}
\frac{\eta_{\bot}}{s} =\frac{8\pi T^2}{3\rho^2}~\label{eq:eta_higha}.
\end{equation}
We see that in this limit the ratio can be made arbitrarily small, with
$\frac{\eta_{\bot}}{s}\rightarrow 0,$ as $T \rightarrow 0$ keeping $\rho$
fixed. \footnote{In this regime $\eta_{\bot} \sim \frac{T^4}{\rho}$ and $s \sim T^2 \rho$ , whereas for the isotropic case ($\rho=0$)  $\eta_{\bot} \sim T^3$  and $s \sim T^3$. Thus we see that for $T \ll \rho$,  $\eta_{\bot}$ is smaller than its value in the isotropic case while $s$ is bigger, resulting in the parametric violation in Eq.~\ref{eq:eta_higha}.}
}
\end{enumerate}

In contrast the $\eta_{xyxy}$ component (which couples to a spin $2$ metric perturbation) was found to be unchanged from its value in the isotropic case, 
\begin{equation}
\label{spin2}
\frac{\eta_{xyxy}} {s}= \frac{1}{4\pi}
\end{equation}
and thus continues to meet the KSS bound.

\section {Derivation of hydrodynamic modes}
In this appendix, we will first show that the Elliptic mode and the Scissor mode
satisfy the equations of superfluid hydrodynamics in the presence of a harmonic
trap. There are viscous corrections to the hydrodynamic equations, but we work
in a limit where viscous corrections are small and therefore the solutions to
the ideal hydrodynamics can be used to calculate the energy loss rate due to
viscosity in a perturbative manner.

\subsection{Equations of superfluid hydrodynamics}
\label{superfl}
Neglecting viscosity, the superfluid equations are given by the conservation laws of entropy, mass
(particle number), momentum and an additional equation for the superfluid
velocity. In the presence of the external potential $\phi(\bfr)$ they are listed below : 
\begin{align}
\label{s_cons}
&\frac{\partial (\rho s)}{\partial t} 
 +  \nabla \cdot{(\rho s \bfv_n)} = 0, \\\,
\label{rho_cons}
&\frac{\partial \rho}{\partial t} 
 +  \nabla \cdot{\bf g} = 0 \, , \\
\label{g_cons}
&\frac{\partial  g_i}{\partial t } 
 + \nabla_j \Pi_{ij} = -n \nabla \phi(\bfr),  \,  \\ 
\label{vs_euler}
&\frac{\partial {\bf v}_s}{\partial t} 
   = - \nabla (\frac{\bfv_s^2}{2}+\frac{\phi(\bfr)}{m}+\frac{\mu(\bfr)}{m}) \,. 
\end{align}
Here $\rho$ is the total mass density (where $\rho_n$ and $\rho_s$ are the
normal and superfluid mass density of the system and the total mass density
$\rho=\rho_n+\rho_s$). We have not written out the dependence of the velocity on
position and time. $\mu(\bfr)$ can be thought of as the local chemical
potential. $n$ (not in the subscript) denotes the total number density (which is related to the total mass density $\rho$ via the relation $\rho=m n$), ${\bf g}$ is the
momentum density, and $\Pi_{ij}$ is the stress tensor, given as follows
\begin{equation}
\begin{split}
\label{g_sfl}
 {\bf g} &= \rho_n{\bf v}_n + \rho_s {\bf v}_s\, , \\
 \Pi_{ij} &= P\delta_{ij} + \rho_n\bfv_{n,i}\bfv_{n,j} 
    + \rho_s \bfv_{s,i}\bfv_{s,j} \;. 
\end{split}
\end{equation}
Let us note that  the equation for energy conservation can be derived from the set of equations above, and is not an additional independent constraint.

Altogether there are $8$ equations above  and they can be solved
for the $8$ independent variables - $6$ components of ($\bf{v}_s,\;\bf{v}_n$)
and $T,\;\mu(\bfr)$. We can then express all thermodynamic variables as functions of
$(T,\;\mu(\bfr))$ like $P(T,\;\mu(\bfr))$, $s(T,\;\mu(\bfr))$ etc. In the trap
geometries we consider, the center of the trap is superfluid and the outer trap
is in the normal phase. The equations for a normal fluid can be obtained by
simply substituting $\rho_s=0$ and ignoring Eq.~\ref{vs_euler}. 

Let us first look at the equilibrium situation $\bf{v_n}=\bf{v_s}=0$ in the
absence of external potential $\phi$. Eqns.~\ref{s_cons},~\ref{rho_cons},
\ref{g_cons},~\ref{vs_euler} are satisfied with $\mu(\bfr)$ and $P$ spatially
constant.  

Before we consider the effects of an external potential let us also note that the pressure and number density in the absence of the trap, which we denote as
$ P_{\phi=0}, n_{\phi=0} $
  respectively,
satisfy the Gibbs-Duhem relation  
\begin{equation}
\label{GD}
\frac{\partial
P_{\phi=0}}{\partial \mu}= n_{\phi=0}.  
\end{equation}
In the presence of the external potential $\phi(\bfr)$ with $\bfv_s=\bfv_n=0$, only Eq.~\ref{g_cons} and
Eq.~\ref{vs_euler} changes.  Eq.~\ref{vs_euler} is satisfied by taking 
\begin{equation}
\label{muans}
\mu(\bfr)=\mu -\phi(\bfr),
\end{equation}
 where $\mu$ is a global constant that determines the
total number of particles in the system. Eq.~\ref{g_cons} in the presence of 
$\phi(\bfr)$ becomes 
\begin{equation}
\partial_i P (\bfr)= - n ~\partial_i \phi(\bfr).
\end{equation}
This is consistent with the replacement $\mu(\bfr) \rightarrow \mu -\phi(\bfr)$
if we take the pressure $P$ at a point $\bfr$ in the presence of the trap to be
equal to $P_{\phi=0}(T, \mu - \phi(\bfr))$  and the number density to be $n_{\phi=0}(T, \mu -\phi(\bfr))$.
This follows from Eq.~(\ref{GD}), since $\partial_i P = -\frac{\partial
P_{\phi =0}}{\partial \mu} \partial_i \phi = -n_{\phi=0} ~\partial_i \phi$. This is also
known as LDA (Local Density Approximation). Generally LDA corresponds to the conditions, 
\begin{equation}
\label{ldac}
f(\mu(\bfr),\;T):= f_{\phi = 0}~\left(\mu - \phi(\bfr),\;T\right)
\end{equation}
where $f$ is $P$, $n$, $\rho$ or $s$.  In all the subsequent discussions, a
subscript $0$ indicates that the conditions for  LDA are valid in equilibrium.
Note that in equilibrium $T$ is a constant. 

\subsection{Scissor mode solution to linear order}
\label{sci}
First we look for solutions of the form  
\begin{equation}
\label{condsm}
\bfv_n = \bfv_s = \bf{v}
\end{equation}
 and $ \nabla
\times \bf{v} =0 $. We restrict ourselves to small velocities and linearize the above
equations. For the scissor mode we see from Eq.~\ref{eq:vprofile} and Eq.~\ref{scissor} that 
$\bfv$ is given by 
\begin{equation}
\label{velsm}
\bfv=\alpha~ e^{i\omega t} (z \hat{x}+ x\hat{z})
\end{equation}
where $\alpha=\alpha_x=\alpha_z$ is a constant. 
We will solve the equations to linear order in $\alpha$. 

Let us first explore Eq.~\ref{vs_euler}.  Out of equilibrium ($\bfv\neq0$),
$\mu(\bfr)$ has an extra correction associated with
$\bfv$,
\begin{equation}
\label{mucorr}
\mu(\bfr)=\mu - \phi(\bfr) + \epsilon(\bfr,t)\;.
\end{equation}
Eq.~\ref{vs_euler} then gives
\begin{equation}
\label{ep}
\epsilon= - \alpha m x z~ i \omega~ e^{i \omega t}.
\end{equation}

Once we are out of equilibrium, we will see that the remaining equations are self consistently solved by letting 
\begin{equation}
\label{ans}
f_{\phi\neq 0}(\mu(\bfr),\;T):= f_{\phi = 0}~\left(\mu - \phi(\bfr) + \epsilon(\bfr,t),\;T\right) 
\end{equation}
where $f$ is $P$, $n$, $\rho$ or $s$.\\
The mass and momentum conservation equations, with the condition Eq.~(\ref{condsm}), give 
\begin{align}
\label{cont_nrfl}
 \frac{\partial\rho}{\partial t} & 
 +  \nabla \cdot\left(\rho{\bfv}\right) = 0 \, , \\
\label{euler}
 \rho \frac{\partial{\bfv}}{\partial t}& +\rho ( \bfv.\nabla) \bfv
 = - \nabla P - n \nabla \phi \, 
\end{align} 
where $\phi(\bfr)$ is the external potential and $\rho$ is the total mass density ($\rho_n + \rho_s $).
Linearizing these equations to order $\alpha$ \footnote{ Note that $\epsilon$ in Eq.~\ref{ep} is of order $\alpha$} using Eq.~\ref{ans} we get,
\begin{align}
\label{m1}
& \frac{\partial{\rho}_0 }{\partial \mu}\frac{\partial\epsilon}{\partial t}  
 +  \nabla \cdot\left(\rho_0{\bf v}\right) = 0 \, , \\
 \label{euler2}
&\rho_0 \frac{\partial{\bf v}}{\partial t}  
 = - \nabla(\frac{\partial{P}_0 }{\partial \mu} \epsilon) -(\frac{\partial{n}_0 }{\partial \mu} \epsilon)  \nabla \phi \, .
\end{align}

Using $ \partial_i \rho_0 = -\frac{\partial{\rho}_0 }{\partial \mu} \partial_i
\phi$ and using the fact that for the modes we consider in this paper $ \nabla.{\bf{v}}=0$ we get
from Eq.~\ref{m1}
\begin{equation}
\label{scsol}
 \frac{\partial\epsilon}{\partial t}  
 -\partial _i \phi ~\bfv_i = 0 \,.
\end{equation}
Plugging in the harmonic potential and the solution  Eq.~\ref{ep}, we find that
the above equation is solved by the Scissor mode which satisfies the condition,
Eq.~\ref{scissor}.  Now taking time derivative of the Euler equation
Eq.~\ref{euler2} and using Eq.~\ref{m1} in the second term on R.H.S of
Eq.~\ref{euler2} and $\frac{\partial{P}_0 }{\partial \mu}= n_0$ (total number
density at equilibrium),
\begin{equation}
\begin{split}
&\rho_0 \frac{\partial^2\bfv_i}{\partial t^2}
=-\partial_i(n_0 \frac{\partial{\epsilon}}{\partial{t}})+\partial_j(n_0 \bfv_{j})\partial_{i} \phi \\ 
&\Rightarrow \rho_0 \frac{\partial^2\bfv_i}{\partial t^2}+ n_0\partial_i( \frac{\partial{\epsilon}}{\partial{t}})
=-\partial_i n_0( \frac{\partial{\epsilon}}{\partial{t}})+\partial_j n_0 \bfv_{j}\partial_{i} \phi \\
&\Rightarrow \rho_0 \frac{\partial^2\bfv_i}{\partial t^2}+n_0\partial_i( \frac{\partial{\epsilon}}{\partial{t}})
=\frac{\partial{n}_0 }{\partial \mu} \partial_i \phi(
\frac{\partial{\epsilon}}{\partial{t}})-\frac{\partial{n}_0 }{\partial \mu}
~\partial_j \phi~ \bfv_{j}~\partial_{i} \phi \;.
\end{split}
\end{equation}
We see from Eq.~\ref{scsol} that the RHS of the above equation vanishes. 
For the scissor mode, it follows  from Eq.~\ref{scissor} and Eq.~\ref{ep} that the LHS also vanishes, and thus the equation is met.

For the time dependent scissor mode, the mass conservation equation is 
\begin{equation}
\label{cont_nrfl2}
 \frac{\partial\rho}{\partial t}  
 + { \nabla}\cdot\left(\rho{\bf v}\right) = 0 \,
\end{equation}
for $\bfv_s=\bfv_n=\bfv$.

Starting with Eq.~\ref{s_cons} and using Eq.~\ref{cont_nrfl2} we get 
\begin{equation}
 \frac{\partial  s}{\partial t} 
 + \bfv\cdot{\bm \nabla} { s } = 0 \,.
\end{equation}

Assuming that the entropy is of the form
$s(\mu - \phi(\bfr) + \epsilon(\bfr,t))$ as given in Eq.~\ref{ans} and linearizing in $\alpha$  we get
\begin{equation} 
 \frac{\partial{s}_0 }{\partial \mu}\frac{\partial\epsilon}{\partial t}  
 - \frac{\partial{s}_0 }{\partial \mu}\partial_i \phi~ \bfv_i = 0 \,. 
\end{equation}
This equation is valid when Eq.~\ref{scsol} is met. 
Hence we find that the ansatz Eq.~\ref{ans} with Eq.~\ref{ep} meets all the equations self consistently.

\subsection{Elliptic mode solution to linear order}
\label{stat}
Next we verify that the Elliptic mode, Eq.\ref{modea}, solves the superfluid
equations to linear order in the velocity. Note that this mode is a stationary
solution ($\omega=0$). Like in the previous case we take $T$ to be a constant
in this mode. Note that in this solution $\bfv_n$ has a non-zero curl, $\nabla
\times \bfv_n\ne 0$, and therefore in the absence of vortices $\bfv_s\neq\bfv_n$. We will denote $\bfv_n=\bfv$ below. 

We start with Eq.~\ref{vs_euler}. Since $\bfv_s=0$ in this mode, we see that this equation is met if 
\begin{equation}
\label{muvar}
\mu(r)=\mu -\phi(r)
\end{equation}
where $\mu$ on the RHS is an $\bf{r}$ independent  constant.

Next, with $\bfv_s=0$ the mass and momentum conservation equations simplify to 
\begin{align}
\label{cont_nrfls}
 \frac{\partial\rho}{\partial t}  
 + {\bm \nabla}\cdot\left(\rho_n{\bf v}\right) =& 0 \, , \\
\label{eulerd}
 \frac{\partial{(\rho_n \bf v_i)}}{\partial t} +{\bm \nabla}_j(\rho_n 
 {\bf v_i} {\bf v_j} )
 =& -{\bm \nabla}_i P - n \nabla_i \phi \, .
\end{align}

The time derivatives in these equations can be dropped. 
The Euler equation, Eq.~\ref{eulerd},  is met to order
$\bfv$ if  $P$ and $n$  take their form in the LDA approximation, Eq.~\ref{ldac}.
We will also assume that the other thermodynamic values, $\rho_n, s$ take this  LDA form and denote them with a subscript $0$. 
Using the fact that $\nabla\cdot \bfv=0$, the other equation, Eq.~\ref{cont_nrfls}, becomes, \begin{equation}
\label{m1s}
 {\bm \nabla}\cdot\left(\rho_{0n}{\bf v}\right) = 0 \,
 \Rightarrow -\frac{\partial{\rho}_{0n} }{\partial \mu} ~\partial_i \phi~ \bfv_i=0
\end{equation}
where we have used the ansatz Eq.~\ref{ldac} for the mass density of the normal component. 
For our mode $ \alpha_x z ~\hat{x} + \alpha_z x ~\hat{z}$ with $ \alpha_{z}=- \frac{\omega_{x}^{2}}{\omega_{z}^{2}} \alpha_{x} $  (see Eq.~\ref{modea}) one can easily check that 
\begin{equation}
\label{condsfa}
\partial_i\phi ~\bfv_i=0,
\end{equation}
 so that this equation is satisfied. \\
 Finally, the entropy conservation equation (after replacing $\rho, s$ by their LDA  values)  becomes 
\begin{equation}
 {\bm \nabla} \cdot{(\rho_{0} s_0 {\bf v})} = 0. \,
\end{equation}
Using the fact that our mode is free of divergence, and $\rho_0 s_0$ is a
function of $\mu - \phi(\bfr)$,   we see that this equation is also met when Eq.~\ref{condsfa} is satisfied. 

It is interesting to note that the fact that the Elliptic mode and the Scissor mode  also solve the equations of one component fluid mechanics in the normal phase. 
Since the temperature is a constant in these modes, and the chemical potential varies as given in Eq.~\ref{muans}, up to possible corrections of order $\epsilon$, Eq.~\ref{mucorr}, 
as one moves from the center of the trap to its edges the ratio $\mu(r)/T$ becomes smaller  and the system will  transit from the superfluid to normal phase. 
The solutions we have found above, for both modes, will continue to hold in such situations as well.

\section{Ideal hydrodynamic modes} 
\label{modes}

In this section we contrast the modes discussed in Sec.~\ref{vprofile} with the
breathing modes discussed in Ref.~\cite{Schafer:2007pr} in normal fluids.

We start with the linearized continuity and Euler equations for a fluid with a
polytropic equation of state, which can be used to derive the following
equation valid for ideal fluid dynamics for the normal
component~\cite{Schafer:2007pr}, 
\begin{equation}
m\frac{\partial^2\vec{v}}{\partial t^2} = 
 -\gamma\left(\vec{\nabla}\cdot\vec{v}\right)
        \left(\vec\nabla \phi(\bfr)\right)
 -\vec{\nabla}\left(\vec{v}\cdot\vec{\nabla} \phi(\bfr)\right)~\label{eq:euler}.
\end{equation}

As shown in Ref.~\cite{Schafer:2007pr} breathing modes can be obtained by
considering a scaling ansatz $v_i=a_ix_i \exp(i\omega t)$ (no sum over $i$).
Substituting in Eq.~\ref{eq:euler} one obtains an eigenequation 
\begin{equation}
\left(2\omega_j^2-\omega^2\right) a_j  + \gamma\omega_j^2\sum_k a_k = 0  .
\end{equation}
This is a simple linear equation of the form $M a=0$. Non-trivial 
solutions correspond to $\det(M)=0$. 

In the case of a trapping potential with axial symmetry,
$\omega_1=\omega_2=\omega_0$, $\omega_3=\lambda\omega_0$, we get $\omega^2 =
2\omega_0^2$ and~\cite{Heiselberg:2004zz,Stringari:2008,Bulgac:2007}
\begin{align}
\label{w_rad}
 \omega^2 =& \omega_0^2\left\{ \gamma+1+\frac{\gamma+2}{2}\lambda^2 
  \right. \\
 & \left. \pm \sqrt{ \frac{(\gamma+2)^2}{4}\lambda^4
           + (\gamma^2-3\gamma-2)\lambda^2 
           + (\gamma+1)^2 }\right\}. \nonumber 
\end{align}

In the unitarity limit ($\gamma=2/3$) and for a very asymmetric trap, 
$\lambda\to 0$, the eigen-frequencies are $\omega^2=2\omega_0^2$ and 
$\omega^2=(10/3)\omega_0^2$. The mode $\omega^2=(10/3)\omega_0^2$ is 
a radial breathing mode with $\vec{a} = (a,a,0)$ and the mode 
$\omega^2=2\omega_0^2$ corresponds to a radial quadrupole $\vec{a} = 
(a,-a,0)$.

Here we consider a different class of modes, with the scaling form
Eq.~\ref{velp} (since $x$ and $z$ are exchanged, they are ``twisted''). The
eigen-equations are now given by Eq.~\ref{eq:twisted_eigen}. It has two
solutions, $\omega=0$ and $\omega= \sqrt{\omega_{x}^2 +\omega_{y}^2}$. 
Hydrodynamic modes can be obtained by considering an ansatz of the form
\begin{equation}
v = e^{i \omega t} (\alpha_{x}~z~ \hat{x} + \alpha_{z}~x ~\hat{z})\;.
\label{velp}
\end{equation}

Substituting Eq.~\ref{velp} in Eq.~\ref{eq:euler} gives the simultaneous
equations
\begin{equation}
\begin{split}
\omega^{2} \alpha_{z} &=  \alpha_{x}~ \omega_{x}^{2} + \alpha_{z}~
\omega_{z}^{2}\\
\omega^{2} \alpha_{x} &=  \alpha_{x}~ \omega_{x}^{2} + \alpha_{z}~
\omega_{z}^{2}\;.
\label{eq:twisted_eigen}
\end{split}
\end{equation}

One mode of interest for us is the $\omega=0$ mode since it has a velocity
profile similar to Fig.~\ref{plates}. This is what we call {the Elliptic mode}.
If $\omega_x=\omega_z$, the mode looks like a rigid body rotation and can not
exhibit viscous damping. For $\omega_x\neq\omega_z$ however we get a non-zero
energy dissipation due to viscosity given by Eq.~\ref{eq:dissipation}. The
second mode of interest for us is what we call the Scissor mode which is well known in literature.  

\section{Anisotropic viscosities in the relaxation time approximation}
\label{microscopic}
In this section, we compute the anisotropic shear viscosities associated with
the motion of a weakly interacting Fermi gas in the presence of an external
potential in the relaxation time approximation~\cite{Ofengeim:2015qxz}. For
this section we explicitly keep $\hbar$ and $c$ in the expressions to ease
comparisons with existing literature.

The Boltzmann equation in the relaxation time approximation is
\begin{equation}
\frac{\partial f(x,p)}{\partial x^\alpha}V_\alpha
+
\frac{\partial f(x,p)}{\partial p^\alpha}(-\nabla_\alpha\phi)
= -\frac{\delta f}{\tau}
\end{equation}
where $f$ is the distribution function, and $\tau$ is the effective 
relaxation time.

In equilibrium, the distribution function of occupied states for a weakly
interacting gas is given by the Fermi-Dirac distribution function
$f_0(x,p)=1/\{\exp[(\epsilon(p)-p\cdot V(x)-\mu)/T(x)]+1\}$, where $\epsilon$,
$p$ represent electron energy and momentum respectively. If a slowly varying local fluid 
velocity $V_\alpha$ ($\alpha=1,\;2,\;3$) is set up in the system, the electron
distribution function is modified. To the lowest order in the derivatives of
$V_\alpha$, we can write   
\begin{equation}
\label{eq:f=f0+df}
f(p) = f_0(\epsilon) + \delta f(p),
\end{equation}
where the non-equilibrium correction $\delta f(p)$ is of the form
where 
\begin{equation}
\label{delfV}
 \delta f(p)=-
 \left( \partDer{f_0}{\mu} \right)v_{\alpha} p_{\beta} 
 C_{\alpha \beta \gamma \delta}(\epsilon)V_{\gamma \delta}\,
\end{equation}
where  $C_{\alpha \beta \gamma \delta}$ is a 4-rank tensor, $\mu$ represents
the electron chemical potential, $v_\alpha=d\epsilon/d{p_\alpha}$ denotes
the electron velocity, and ${V}_{\alpha \beta }$ is proportional to the
derivative of the macroscopic fluid velocity defined as follows
\begin{equation}
\label{eq:Vdef}
V_{\alpha\beta} = \frac{1}{2} \left(
\partDer{V_{\alpha}}{x_{\beta}} + \partDer{V_{\beta}}{x_{\alpha}}
\right),
\end{equation}

Similarly, in the presence of a slowly varying external potential $\phi$,
Eq.~\ref{eq:f=f0+df} holds with 
\begin{equation}
\label{delfphi}
 \delta f(p)=
 - \left( \partDer{f_0}{\mu} \right)v_{\alpha} D_{\alpha \gamma}(\epsilon)
 \partial_{\gamma} {\phi}\;.
\end{equation}
Here we consider both $\partial_\alpha\phi$ and $V_{\alpha\beta}$ non-zero, and
hence $\delta f$ is the sum of Eq.~\ref{delfV} and Eq.~\ref{delfphi}.   
After canceling out the terms proportional to $D$ (which are related to
conductivity) the linearized Boltzmann equation within the relaxation time approximation of
the collision integral takes the form 
\begin{equation}
\label{eq:boltzmanelec}
\left( \partDer{f_0}{\mu} \right)\left( v_{\alpha} p_{\beta}
\partDer{V_{\alpha}}{x_{\beta}} - \frac{1}{3}v_{\alpha}p_{\alpha}
\diverg {V} \right) = -\frac{\delta f}{\tau} +\, (
\nabla \phi) \cdot \partDer{\delta f}{{p}}\;,
\end{equation}
in analogy with Eq.~$2$ of \cite{Ofengeim:2015qxz} for the magnetic field case,
\begin{equation}
\left( \partDer{f_0}{\mu} \right)\left( v_{\alpha} p_{\beta}
\partDer{V_{\alpha}}{x_{\beta}} - \frac{1}{3}v_{\alpha}p_{\alpha}
\diverg {V} \right) = -\frac{\delta f}{\tau} + \frac{e}{c}\, (
{v} \times {B})\cdot 
\partDer{\delta f}{{p}}\;.~\label{eq:BoltzmannB}
\end{equation}

For ease of calculation, let us decompose the $\nabla \phi$ term on the R.H.S
of Boltzmann equation as 
\begin{equation}
\label{decom}
\nabla \phi= \hat{p}( \hat{p}.\nabla \phi) +(\nabla \phi- \hat{p}( \hat{p}.\nabla \phi))=\hat{p}( \hat{p}.\nabla \phi) +\hat{p} \times (\nabla \phi\times \hat{p})
\end{equation}

In what follows, it is useful to define a basis $\xi^{'}$  for the 8
dimensional non-commutative algebra for the 4-rank tensor $C_{\gamma \delta \mu
\nu}$ built out of the Kroenecker delta, Levi-civita and the components of the
unit vector along the direction $\nabla \phi \times \hat{p} $ denoted by
$\hat{b}$.

The basis $\xi'_1 - \xi'_8$ is defined as
\begin{equation}
\label{basis1}
\begin{aligned}
&\xi'_{1_{\alpha\beta\gamma\delta}}=\delta _{\alpha \gamma } \delta _{\beta \delta }+\delta
   _{\alpha \delta } \delta _{\beta \gamma }\\
&\xi'_{2_{\alpha\beta\gamma\delta}}=\delta _{\alpha \beta } \delta _{\gamma \delta } \\
&\xi'_{3_{\alpha\beta\gamma\delta}}=\hat{b}_{\alpha } \hat{b}_{\delta } \delta _{\beta \gamma
   }+\hat{b}_{\alpha } \hat{b}_{\gamma } \delta _{\beta \delta
   }+\delta _{\alpha \gamma } \hat{b}_{\beta } \hat{b}_{\delta
   }+\delta _{\alpha \delta } \hat{b}_{\beta } \hat{b}_{\gamma }\\
&\xi'_{4_{\alpha\beta\gamma\delta}}=\delta _{\alpha \beta } \hat{b}_{\gamma } \hat{b}_{\delta }\\
&\xi'_{5_{\alpha\beta\gamma\delta}}=\hat{b}_{\beta } \hat{b}_{\delta } \delta _{\gamma \delta }\\
&\xi'_{6_{\alpha\beta\gamma\delta}}=\hat{b}_{\alpha } \hat{b}_{\beta } \hat{b}_{\gamma } \hat{b}_{\delta }\\
&\xi'_{7_{\alpha\beta\gamma\delta}}=\delta _{\alpha \gamma } \hat{b}_{\beta \delta }+\hat{b}_{\alpha
   \gamma } \delta _{\beta \delta }+\delta _{\alpha
   \delta } \hat{b}_{\beta \gamma }+\hat{b}_{\alpha \delta } \delta
   _{\beta \gamma }\\
&\xi'_{8_{\alpha\beta\gamma\delta}}=\hat{b}_{\alpha } \hat{b}_{\beta \gamma } \hat{b}_{\delta }+\hat{b}_{\alpha }
   \hat{b}_{\beta \delta } \hat{b}_{\gamma }+\hat{b}_{\alpha \gamma }
   \hat{b}_{\beta } \hat{b}_{\delta }+\hat{b}_{\alpha \delta } \hat{b}_{\beta }
   \hat{b}_{\gamma }
\end{aligned}
\end{equation}
Let us now simplify the L.H.S of Eq.~\ref{eq:boltzmanelec}
\begin{equation}
\begin{split}
&\left( \partDer{f_0}{\mu} \right)\left( v_{\alpha} p_{\beta}
\partDer{V_{\alpha}}{x_{\beta}} - \frac{1}{3}v_{\alpha}p_{\alpha}
\diverg \bm{V} \right)\\
&=\left( \partDer{f_0}{\mu} \right) v_\alpha p_\beta V_{\mu \nu}\frac{1}{2} \left(\xi^{'}_{1_{\alpha \beta \mu \nu}} - \frac{2}{3} \xi^{'}_{2_{\alpha \beta \mu \nu}}\right)
\end{split}
\end{equation}

Similarly the R.H.S of Eq.~\ref{eq:boltzmanelec} can be simplified as follows-
\begin{equation}
\begin{split}
{\rm{R.H.S}}&=-\frac{\delta f}{\tau} + \, (
\nabla \phi)_\alpha
\partDer{\delta f}{\bm{p}_\alpha}=-\frac{\delta f}{\tau}+(\hat{p}( \hat{p}.\nabla \phi) +\hat{p} \times (\nabla \phi\times \hat{p}))_\alpha \partDer{\delta f}{\bm{p}_\alpha} \\
&=-\delta f \left(\frac{1}{\tau}-\left(\frac{ \hat{p}.\nabla \phi }{p}\right)\right)  -\left( \hat{p} \times (\nabla \phi\times \hat{p})\right)_\alpha v_a C_{a \alpha \gamma \delta}V_{\gamma \delta}\left( \partDer{f_0}{\mu}\right)
\end{split}
\end{equation}

Taking $\tau$ to the L.H.S we get 
\begin{equation}
\begin{split}
\tau {\rm{L.H.S}}
&= -\delta f \left(1-{\tau}\left(\frac{ \hat{p}.\nabla \phi }{p}\right)\right)  -\tau \left( \hat{p} \times (\nabla \phi\times \hat{p})\right)_\alpha v_a C_{a \alpha \gamma \delta}V_{\gamma \delta}\left( \partDer{f_0}{\mu}\right)\\
&=v_\alpha p_\beta V_{rs}\left( \partDer{f_0}{\mu}\right)\left( C_{\alpha \beta rs} \left(1-\tau \frac{ \hat{p}.\nabla \phi }{p}\right)- \frac{\tau b}{ p}\epsilon_{\theta \beta \gamma}\hat{b}_\gamma C_{\alpha \theta rs} \right)\\
\end{split}
\end{equation}
where $b$ denotes the magnitude of the vector $\nabla \phi \times \hat{p}$.

Let $a= \left(1-\tau\frac{ \hat{p}.\nabla \phi }{p}\right)$ and $ x= \frac{
\tau b}{ p}$. If we denote the angle between $\nabla \phi$ and $\hat{p}$ as
$\theta$, then $ a= 1-\frac{\E \tau}{p} \cos \theta$ and 
$x= \frac{\tau\E}{ p} \sin \theta$.

Hence we get
\begin{equation}
\tau {\rm{L.H.S}} = v_\alpha p_\beta V_{rs} \left( \partDer{f_0}{\mu}\right) \left(a C_{\alpha \beta rs} -x \epsilon_{\theta \beta \gamma} \hat{b}_\gamma C_{\alpha \theta rs}\right)
\end{equation}

Symmetrizing in $\alpha$ and $\beta$, we get 
\begin{equation}
\begin{split}
\tau {\rm{L.H.S}} &= v_\alpha p_\beta V_{rs} \left( \partDer{f_0}{\mu}\right) \left(a\frac{ C_{\alpha \beta rs} + C_{\beta \alpha rs}}{2}-x \frac{\epsilon_{\theta \beta \gamma} \hat{b}_\gamma C_{\alpha \theta rs}+\epsilon_{\theta \alpha \gamma} \hat{b}_\gamma C_{\beta \theta rs}}{2}\right)\\
&=v_\alpha p_\beta V_{rs}\frac{1}{2} C_{\gamma \delta rs} \left( \partDer{f_0}{\mu}\right) \left(a \delta_{\alpha \gamma} \delta _{\beta \delta} +a \delta_{\beta \gamma}\delta_{\alpha \delta}+x (b_{ \beta\delta} \delta_{\gamma \alpha}+ b_{ \alpha\delta}\delta_{\gamma \beta}) \right)\\
\end{split}
\end{equation}

Subtracting the trace in $\alpha \beta$, we get
\begin{equation}
\begin{split}
\tau {\rm{L.H.S}} &=v_\alpha p_\beta V_{rs}\frac{1}{2} C_{\gamma \delta rs} \left( \partDer{f_0}{\mu}\right) \left(a \delta_{\alpha \gamma} \delta _{\beta \delta} +a \delta_{\beta \gamma}\delta_{\alpha \delta} - \frac{2}{3}a\delta_{\gamma \delta}\delta_{\alpha \beta}+x (b_{ \beta\delta} \delta_{\gamma \alpha}+ b_{ \alpha\delta}\delta_{\gamma \beta} + b_{\alpha \gamma} \delta_{\beta \delta} + b_{\beta \gamma} \delta_{\alpha \delta}) \right)\nonumber\\
&=v_\alpha p_\beta V_{rs}\frac{1}{2} C_{\gamma \delta rs} \left( \partDer{f_0}{\mu}\right) \left(a \xi'_1- \frac{2}{3}a\xi'_2+x \xi'_7 \right)_{\alpha \beta \gamma \delta}\nonumber\\
\end{split}
\end{equation}

Now combining L.H.S  and R.H.S we finally get
\begin{equation}
\begin{split}
&\tau \left( \partDer{f_0}{\mu} \right) v_\alpha p_\beta V_{\mu \nu}\frac{1}{2} \left(\xi^{'}_{1_{\alpha \beta \mu \nu}} - \frac{2}{3} \xi^{'}_{2_{\alpha \beta \mu \nu}}\right)=v_\alpha p_\beta V_{rs}\frac{1}{2} C_{\gamma \delta rs} \left( \partDer{f_0}{\mu}\right) \left(a \xi'_1- \frac{2}{3}a\xi'_2+x \xi'_7 \right)_{\alpha \beta \gamma \delta}\nonumber\\
&\Rightarrow \tau  \left(\xi^{'}_{1_{\alpha \beta \mu \nu}} - \frac{2}{3} \xi^{'}_{2_{\alpha \beta \mu \nu}}\right)=   \left(a \xi'_1- \frac{2}{3}a\xi'_2+x \xi'_7 \right)_{\alpha \beta \gamma \delta} C_{\gamma \delta \mu \nu}\nonumber\\
\end{split}
\end{equation}

Writing $ C_{\gamma \delta \mu \nu}=\left(\sum_{i=1}^8 c_i
\xi^{'}_{i\,\gamma\delta \mu \nu}\right)$ we can now solve for the coefficients
\begin{equation}
\label{sol1}
\begin{aligned}
&c_1 = \frac{ a \tau}{2(a^2 + 4 x^2)}, ~c_2=- \frac{\tau (a^2-2 x^2)}{3 a (a^2+4 x^2)},~ c_3= \frac{3 a \tau x^2}{2 (a^2+x^2)(a^2+4x^2)},
c_4=c_5=-\frac{2 \tau x^2}{a(a^2+4x^2)},\\
 &c_6= \frac{6 \tau x^4}{a(a^2+x^2)(a^2+4x^2)},~c_7=-\frac{\tau x}{2(a^2+4x^2)},~c_8= - \frac{ 3 \tau x^3}{2(a^2+x^2)(a^2+4x^2)}
\end{aligned}
\end{equation}

The viscosity tensor is given as
\begin{equation}
\label{eq:sigmaDef}
\eta_{\alpha\beta a b} = -
\frac{2}{(2\pi \hbar)^3}
\int {\rm d^3} \bm{p}\, 
\left( \partDer{f_0}{\mu} \right) 
v_{\alpha} p_{\beta}  v_{\gamma} p_{\delta}\,
\left(\sum_{i=1}^8 c_i \xi^{'}_{i\,\gamma\delta a b}\right).
\end{equation}

It is convenient to decompose the tensor $\eta_{\alpha\beta a b}$ in to $5$
irreducible components corresponding to $5$ tensors $M_{i\,\alpha\beta a b}$
($i=0,\cdot\cdot4$) in a system with a special direction
$\hat{E}=\nabla\phi/|\nabla\phi|$ and reflection symmetry. 
\begin{equation}
\eta_{\alpha\beta \gamma \delta} = \sum_{i=0}^4 \eta_i 
M_{i\,\alpha\beta\gamma \delta}\;.
\end{equation}

The tensors $M_i$ are
\begin{equation}
\begin{aligned}
&M_0=3 \xi_6 - \xi_4 - \xi_5 +\frac{\xi_2}{3}\\
&M_1=\xi_1- \xi_2 -\xi_3 +\xi_4+ \xi_5+ \xi_6 \\
&M_2=\xi_3 - 4\xi_6 \\
&M_3=-\frac{1}{2}(\xi_7 -\xi_8 )\\
&M_4=-\xi_8
~\label{eq:projections}
\end{aligned}
\end{equation}
where the basis $\xi_1 - \xi_8$ is defined as
\begin{equation}
\label{basis2}
\begin{aligned}
&\xi_{1_{\alpha\beta\gamma\delta}}=\delta _{\alpha \gamma } \delta _{\beta \delta }+\delta
   _{\alpha \delta } \delta _{\beta \gamma }\\
&\xi_{2_{\alpha\beta\gamma\delta}}=\delta _{\alpha \beta } \delta _{\gamma \delta } \\
&\xi_{3_{\alpha\beta\gamma\delta}}=\hat{E}_{\alpha } \hat{E}_{\delta } \delta _{\beta \gamma
   }+\hat{E}_{\alpha } \hat{E}_{\gamma } \delta _{\beta \delta
   }+\delta _{\alpha \gamma } \hat{E}_{\beta } \hat{E}_{\delta
   }+\delta _{\alpha \delta } \hat{E}_{\beta } \hat{E}_{\gamma }\\
&\xi_{4_{\alpha\beta\gamma\delta}}=\delta _{\alpha \beta } \hat{E}_{\gamma } \hat{E}_{\delta }\\
&\xi_{5_{\alpha\beta\gamma\delta}}=\hat{E}_{\beta } \hat{E}_{\delta } \delta _{\gamma \delta }\\
&\xi_{6_{\alpha\beta\gamma\delta}}=\hat{E}_{\alpha } \hat{E}_{\beta } \hat{E}_{\gamma } \hat{E}_{\delta }\\
&\xi_{7_{\alpha\beta\gamma\delta}}=\delta _{\alpha \gamma } \hat{E}_{\beta \delta }+\hat{E}_{\alpha
   \gamma } \delta _{\beta \delta }+\delta _{\alpha
   \delta } \hat{E}_{\beta \gamma }+\hat{E}_{\alpha \delta } \delta
   _{\beta \gamma }\\
&\xi_{8_{\alpha\beta\gamma\delta}}=\hat{E}_{\alpha } \hat{E}_{\beta \gamma } \hat{E}_{\delta }+\hat{E}_{\alpha }
   \hat{E}_{\beta \delta } \hat{E}_{\gamma }+\hat{E}_{\alpha \gamma }
   \hat{E}_{\beta } \hat{E}_{\delta }+\hat{E}_{\alpha \delta } \hat{E}_{\beta }
   \hat{E}_{\gamma }\;,
\end{aligned}
\end{equation}
where $\hat{E}$ is the unit vector along the gradient of the potential.

The components $\eta_i$ can be extracted by projecting onto $M_i$ and
performing the three dimensional momentum integral in Eq.~\ref{eq:sigmaDef}. For arbitrarily
large $\frac{|\tau\nabla\phi|}{k_F}$ the momentum integrals can not be performed
analytically in general. However, we are interested in
$\frac{|\tau\nabla\phi|}{k_F}\lesssim 1$, where the corrections to isotropy
just start to become important. Then one can expand in ${|\tau\nabla\phi|}$ and
perform the angular integrals to obtain, 
\begin{equation}
\begin{split}
\eta_0 &= \eta(0)[1-\frac{31}{21}\tau^2(\E)^2\frac{I_2}{I_1}+\calO((\tau\E)^4)]\\
\eta_1 &= \eta(0)[1-\frac{13}{7}\tau^2(\E)^2\frac{I_2}{I_1}+\calO((\tau\E)^4)]\\
\eta_2 &= \eta(0)[1-\frac{11}{7}\tau^2(\E)^2\frac{I_2}{I_1}+\calO((\tau\E)^4)]\\
\eta_3 &= 0, ~\eta_4 = 0\;,
\label{etaft}
\end{split}
\end{equation}
where
\begin{equation}
\eta(0) = \int p^6 dp~\frac{\tau}{15 \pi^2 m^2  {\hbar}^3} \left( \partDer{f_0}{\mu} \right)
\end{equation}
is the shear viscosity in the absence of $\nabla\phi$, and 
$I_1$ and $I_2$ are.
\begin{equation}
I_1~=\int p^6 dp\left( \partDer{f_0}{\mu} \right),\\
I_2~=\int p^4 dp\left( \partDer{f_0}{\mu} \right)
\end{equation}

In particular, in the degenerate limit ($T\ll\mu$) 
\begin{equation}
\left( \partDer{f_0}{\mu} \right)\approx \delta(\frac{p^2}{2m}-\mu)\;,
\end{equation}
and $\frac{I_1}{I_2}\approx \frac{1}{k_F^2}$ where $k_F=(3\pi^2n)^{1/3}$ as
before.

We can write Eq.~\ref{etaft} in the form Eq.~\ref{eq:anisotropy_corrections0}
by relating the relaxation time $\tau$ to the mean free path $\lambda$
\begin{equation}
\frac{\tau}{k_F}=\frac{\tau}{k_F}\frac{E_F}{E_F} = \frac{\lambda}{2E_F}
\end{equation}
where we have used $E_F/k_F=v_F/2$, and $\tau v_F=\lambda$ is the mean free
path.

This gives,
\begin{equation}
\begin{split}
\eta_0 &=
\eta(0)[1-\frac{31}{84}\frac{\lambda^2(\E)^2}{\mu^2}+\calO((\tau\E)^4)]\\
\eta_1 &=
\eta(0)[1-\frac{13}{28}\frac{\lambda^2(\E)^2}{\mu^2}+\calO((\tau\E)^4)]\\
\eta_2 &=
\eta(0)[1-\frac{11}{28}\frac{\lambda^2(\E)^2}{\mu^2}+\calO((\tau\E)^4)]\\
\eta_3 &= 0, ~\eta_4 = 0\;,
\label{eq:etaoflambda}
\end{split}
\end{equation}
where 
\begin{equation}
\eta(0) = \frac{ (2 m \mu)^\frac{5}{2}\tau}{15\pi^2\hbar^3 m}\;,
~\label{eq:eta0degenerate}
\end{equation}
in the degenerate limit.

Eq.~\ref{eq:etaoflambda} gives an explicit result of the calculation in the relaxation time
approximation which shows that the correction to the viscosity has the form
Eq.~\ref{eq:anisotropy_corrections0}.  Hearteningly, the sign of $c_{(i)}$ is
negative, meaning that the viscosity is reduced due to the external potential,
a feature found is strongly coupled theories where a quasi-particle description
is not possible and hence the Boltzmann transport equation can not be used to
calculate the viscosity.

Interestingly, in the degenerate limit it is possible to do the momentum
integrals analytically for general $\nabla\phi$. Using $\left(
\partDer{f_0}{\mu} \right)= \delta(\frac{p^2}{2m}-\mu)$, we get (here $x =
\frac{\E\tau}{\sqrt{2m \mu}}$ )
\begin{equation}
\begin{aligned}
\eta_0=&\frac{(2 m \mu)^\frac{5}{2} \tau}{96 m \hbar^3 \pi ^2 x^5   \sqrt{3~ x ^2+1}}\Bigg[-8 \sqrt{3 (x ^2+1} \left(5 x^4+18 x ^2+9\right) \tanh
   ^{-1}(x )\Bigg.\\
&\Bigg.-24~ x  \sqrt{3 x^2+1}
   \left(5 x^2+3\right)-6 \left(8
  x^4+11 x^2+3\right)\Bigg.\\
&\Bigg.   \log \left(\frac{x  \left(7 x
   -4 \sqrt{3x^2+1}\right)+1}{x  \left(4 \sqrt{3 x^2+1}+7
   x \right)+1}\right)\Bigg]
\end{aligned}
\end{equation}

\begin{equation}
\begin{aligned}
\eta_1=&\frac{ (2 m \mu)^\frac{5}{2} \tau}{96m \hbar^3 \pi ^2 x^5   \sqrt{3~ x ^2+1}} \Bigg[-4 x \left(x ^2+3\right)
   \sqrt{3 x^2+1}\Bigg.\\
&\Bigg.+4 \sqrt{3
   x^2+1} \left(x^4-6
  x^2-3\right) \tanh ^{-1}(x)-(3 +4 x ^4+9 x ^2)\Bigg.\\
& \Bigg.\log \left(\frac{x  \left(7 x
   -4 \sqrt{3 x ^2+1}\right)+1}{x  \left(4 \sqrt{3 x ^2+1}+7
   x \right)+1}\right)\Bigg]  
\end{aligned}
\end{equation}

\begin{equation}
\begin{aligned}
\eta_2=&\frac{(2 m \mu)^\frac{5}{2} \tau}{48 m \hbar^3 \pi ^2 x^5   \sqrt{3~ x ^2+1}} \Bigg[8 x  \sqrt{3 x ^2+1} \left(4
  x ^2+3\right)\Bigg.\\
&\Bigg.+4 \sqrt{3
  x ^2+1} \left(x ^4+6
  x^2+3\right) \tanh ^{-1}\left(\frac{2
  x }{x ^2+1}\right)+(6+13 x ^4+21x ^2)\Bigg.\\
& \Bigg.\log \left(\frac{x \left(7 x
   -4 \sqrt{3 x ^2+1}\right)+1}{x \left(4 \sqrt{3 x^2+1}+7
   x \right)+1}\right)\Bigg]  \\
  & \eta_3=0\\
  & \eta_4=0
\end{aligned}
\end{equation}

Expanding in small $x$ we obtain, 
\begin{equation}
\begin{split}
\eta_0 &= \frac{ (2 m \mu)^\frac{5}{2}\tau}{15\pi^2\hbar^3 m}
\left(1-\frac{31 \tau^2 \E^2}{42 m\mu} + {\cal{O}}[(\frac{ \tau \E}
{ \sqrt{2 m\mu}})^{4}]\right), \quad \eta_1 =
\frac{ (2 m \mu)^\frac{5}{2}\tau}{15\pi^2\hbar^3 m}
\left(1-\frac{13 \tau^2 \E^2  }{14 m\mu} + {\cal{O}}[(\frac{\tau \E} 
{ \sqrt{2 m\mu}})^4]\right),\nonumber
\\
 \quad \eta_2 &= \frac{ (2 m \mu)^\frac{5}{2}\tau}{15\pi^2\hbar^3 m}
 \left(1-\frac{11 \tau^2 \E^2}{14 m\mu} + {\cal{O}}[(\frac{\tau \E} 
 { \sqrt{2 m\mu}})^4]\right),
\nonumber \\
\eta_3 &= 0, \qquad \eta_4 =
0.\nonumber \label{eq:eta01234}
\end{split}
\end{equation}

\bibliographystyle{ieeetr}
\bibliography{ucver1}

\end{document}